\documentclass[a4paper,11pt]{article}
\pdfoutput=1 % if your are submitting a pdflatex (i.e. if you have
\usepackage{jheppub} % for details on the use of the package, please
\usepackage[T1]{fontenc} % if needed

\usepackage{graphicx}
\usepackage{amsmath,amsfonts,amsthm,amssymb,bbm}
\usepackage[utf8]{inputenc}
\usepackage[usenames]{color}

\definecolor{darkgreen}{rgb}{0.2,0.6,0}
\definecolor{lightblue}{rgb}{0,0.5,0.8}
\definecolor{lightred}{rgb}{0.8,0.2,0.2}
\definecolor{darkorange}{rgb}{1,0.549,0}

\newcommand{\gb}{\bar{g}}
\newcommand{\ub}{\bar{u}}
\newcommand{\Db}{\bar{D}}
\newcommand{\Rb}{\bar{R}}
\newcommand{\hh}{\widehat{h}}

\newcommand{\cD}{{\mathcal D}}

\newcommand{\cN}{{\mathcal N}}
\newcommand{\cO}{{\mathcal O}}
\newcommand{\cR}{{\mathcal R}}

\newcommand{\p}{{\partial}}

\graphicspath{{figures/}}

\newcommand{\be}{\begin{equation}}
\newcommand{\ee}{\end{equation}}
\newcommand{\ba}{\begin{eqnarray}}
\newcommand{\ea}{\end{eqnarray}}

\title{\boldmath On the scaling of composite operators\\ in Asymptotic Safety}

\author[a]{W.\ Houthoff,}
\author[b]{A.\ Kurov,}
\author[a,1]{F.\ Saueressig\note{Corresponding author.}}

\affiliation[a]{Institute for Mathematics, Astrophysics and Particle Physics (IMAPP), \\ Radboud University, Heyendaalseweg 135, 6525 AJ Nijmegen, The Netherlands}
\affiliation[b]{Theory Department, Lebedev Physics Institute, \\ Leninsky Prospect 53, Moscow 119991, Russia}

\emailAdd{wbhouthoff@gmail.com}
\emailAdd{kurov.aleksandr@physics.msu.ru}
\emailAdd{f.saueressig@science.ru.nl}

\abstract{The Asymptotic Safety hypothesis states that the high-energy completion of gravity is provided by an interacting renormalization group fixed point. This implies non-trivial quantum corrections to the scaling dimensions of operators and correlation functions which are characteristic for the corresponding universality class. We use the composite operator formalism for the effective average action to derive an analytic expression for the scaling dimension of an infinite family of geometric operators $\int d^dx \sqrt{g} R^n$. We demonstrate that the anomalous dimensions interpolate continuously between their fixed point value and zero when evaluated along renormalization group trajectories approximating classical general relativity at low energy. Thus classical geometry emerges when quantum fluctuations are integrated out. We also compare our results to the stability properties of the interacting renormalization group fixed point projected to 
	$f(R)$-gravity, showing that the  composite operator formalism in the single-operator approximation cannot be used to reliably determine the number of relevant parameters of the theory.}

\begin{document} 
\maketitle
\flushbottom

\section{Introduction}
Quantum fluctuations of spacetime play a central role in understanding physics at trans-Planckian energy $E \gtrsim E_{\rm Planck} \simeq 10^{18}$ GeV. It is expected that a description of spacetime in this domain requires a yet to be completed theory of quantum gravity. An interesting scenario, which may allow to reconcile gravity with the laws of a relativistic quantum field theory, arises when the quantum fluctuations balance the canonical mass-dimensions of the coupling constants. In this way one may attain a theory where the dimensionless couplings attain constant and finite values at trans-Planckian energies. As a consequence physical quantities like scattering amplitudes would be free from unphysical ultraviolet (UV) divergences. This scenario has been termed Asymptotic Safety \cite{Weinberg:1976xy,Weinberg:1980gg}, also see \cite{Percacci:2017fkn,Reuter:2019byg} for pedagogical introductions. 

At the technical level Asymptotic Safety is realized through a non-Gaussian fixed point (NGFP) of the gravitational renormalization group (RG) flow. Starting from the seminal work \cite{Reuter:1996cp}, functional renormalization group equations (FRGEs) \cite{Wetterich:1992yh,Morris:1993qb,Reuter:1996cp} have produced substantial evidence that gravity indeed possesses a suitable NGFP. This comprises the demonstration that the NGFP seen in four spacetime dimensions is the analytic continuation of the perturbative fixed point seen in $2+\epsilon$ spacetime dimensions \cite{Reuter:2001ag}, studies of the gravitational RG flow based on the  Einstein-Hilbert truncation
\cite{Souma:1999at,Falkenberg:1996bq,Reuter:2001ag,Lauscher:2001ya,Litim:2003vp,
	Bonanno:2004sy,Eichhorn:2009ah,Manrique:2009uh,
	Eichhorn:2010tb,Groh:2010ta,Manrique:2010am,Christiansen:2012rx,Codello:2013fpa,Benedetti:2013jk,
	Becker:2014qya,
	Falls:2015qga,
	Gies:2015tca,
	Pagani:2016dof,Falls:2017cze,
	Knorr:2017fus}, 
and its extension by higher-derivative and higher-order curvature terms \cite{Lauscher:2002sq,Reuter:2002kd,
	Codello:2006in,Codello:2007bd,Machado:2007ea,
	Niedermaier:2009zz,Benedetti:2009rx,Benedetti:2009gn,Benedetti:2009iq,Benedetti:2010nr,Rechenberger:2012pm,
	Ohta:2013uca,Falls:2013bv,Benedetti:2013jk,
	Falls:2014tra,
	Eichhorn:2015bna,Ohta:2015efa,
	Falls:2016wsa,Falls:2016msz,Christiansen:2016sjn,
	Gonzalez-Martin:2017gza,Becker:2017tcx} also including the notorious Goroff-Sagnotti two-loop counterterm \cite{Gies:2016con}, the construction of fixed functions including an infinite number of coupling constants \cite{Reuter:2008qx,Benedetti:2012dx,Demmel:2012ub,Dietz:2012ic,Bridle:2013sra,Dietz:2013sba,Demmel:2014sga,Demmel:2014hla,Demmel:2015oqa,Ohta:2015fcu,Labus:2016lkh,Dietz:2016gzg,Knorr:2017mhu,Falls:2017lst}, the computation of correlation functions retaining arbitrary momentum dependence \cite{Christiansen:2014raa, Meibohm:2015twa, Christiansen:2015rva, Denz:2016qks, Christiansen:2017cxa, Christiansen:2017bsy, Eichhorn:2018akn, Eichhorn:2018ydy}, and the form factor program recently initiated in \cite{Bosma:2019aiu,Knorr:2019atm}.
As a key result, these works indicate that the NGFP comes with substantial predictive power.\footnote{Notably, NGFPs similar to the one encountered in the case of gravity also appear in many phenomenologically interesting gravity-matter systems, see \cite{Dona:2013qba,Christiansen:2017cxa,Biemans:2017zca,Eichhorn:2018whv,Alkofer:2018fxj,Eichhorn:2018akn,Eichhorn:2018ydy,Pawlowski:2018ixd,deBrito:2019epw,Burger:2019upn} for recent results and \cite{Eichhorn:2018yfc} for a review. For recent discussions related to fundamental aspects of Asymptotic Safety we refer to \cite{Morris:2018zgy,Baldazzi:2018mtl,Baldazzi:2019kim,Maas:2019eux}.} 

Complementary, Monte Carlo simulations of the gravitational partition function within the Causal Dynamical Triangulation program \cite{Ambjorn:2012jv,Loll:2019rdj} have provided strong evidence for the existence of second order phase transitions \cite{Ambjorn:2011cg,Ambjorn:2012ij,Ambjorn:2019pkp,Ambjorn:2019lrm}. It is tempting to speculate that these are the imprints of the NGFP seen by the functional renormalization group tailored to foliated spacetimes \cite{Manrique:2011jc,Rechenberger:2012dt,Biemans:2016rvp,Biemans:2017zca,Houthoff:2017oam,Knorr:2018fdu,Eichhorn:2019ybe}.

The existence of a NGFP controlling the high-energy behavior of gravity (or any other completion of the gravitational force laws at microscopic scales) raises the intriguing question how to characterize the properties of spacetime in the quantum regime. A first step in this direction has been based on generalized dimensions including the spectral dimension, Hausdorff dimension, or walk dimension \cite{Ambjorn:2005db,Lauscher:2005qz,Reuter:2012xf}. By now, the spectral dimension $d_s$ associated with the short-distance properties of spacetime has been computed in many quantum gravity programs \cite{Carlip:2019onx} with the rather spectacular outcome that even vastly different approaches coincide in the prediction that $d_s = 2$ on microscopic scales. A more refined characterization could then be based on the anomalous scaling dimension of geometric operators comprising for instance, the volumes of spacetime, volumes of surfaces embedded into spacetime, the geodesic length, or correlation functions of fields separated by a fixed geodesic distance \cite{Pagani:2016dof,Pagani:2017tdr,Becker:2018quq,Becker:2019tlf}.  For instance \cite{Pagani:2016dof} computed the anomalous dimension $\gamma_0$ associated with the $d$-dimensional volume operator $\cO_0 = \int d^dx \sqrt{g}$. At the NGFP in four dimensions this anomalous dimension turned out to be $\gamma_0^*|_{d=4} = 3.986$ which was taken as a pointer that ``spacetime could me much more empty than one would naively expect'' \cite{Pagani:2016dof}. Recently, a similar analysis for Stelle gravity has been reported in \cite{Becker:2019fhi}. 

The goal of the present work is to extend the computation of the anomalous dimension of the volume operator to an infinite family of geometric operators given by the integral of $n$ powers of the Ricci scalar $R$
\be\label{C}
\cO_n = \int d^dx \, \sqrt{g} \, R^n \, , \qquad n \in \mathbb{N} \, . 
\ee
These operators can then either be defined with the integral covering the entire spacetime or a local patch. In the later case, diffeomorphism invariance requires adding suitable boundary terms akin to the Gibbons-Hawking term which are not important for the present discussion though.

Our computation of the anomalous scaling dimensions follows the strategy advocated in \cite{Pagani:2016dof} amd utilizes the composite operator formalism based on the effective average action \cite{Pawlowski:2005xe,Igarashi:2009tj,Pagani:2016pad}\footnote{For earlier works also see \cite{Ellwanger:1994iz,DAttanasio:1996tzp,Litim:1998qi}.}, cf.\ eq.\ \eqref{CompositeMaster}.
The key strength of the formalism is that it allows to compute the scaling of geometric operators which are not part of the effective average action. Practically, we then approximate the (regulated) propagators by the Einstein-Hilbert result \cite{Reuter:1996cp,Percacci:2017fkn,Reuter:2019byg} and chose the geometric operators $\cO_n$ \eqref{C}. Our main result is the general formula \eqref{mainresult} giving a closed expression for the anomalous scaling in any dimension $d$. Notably, this is the first time that information on an infinite number of scaling exponents is given.

Our work comes with a three-fold motivation. First, the anomalous dimensions can be used to characterize the universality class associated with the NGFP in terms of properties of geometric operators. In particular, given the anomalous dimension associated with two (or more) geometric operators in the family \eqref{C} allows to eliminate the coarse graining scale $k$. Suppose that $\cO_n$ scales according to $\cO_n \propto k^{d_n + \gamma_n}$ where $d_n$ and $\gamma_n$ are the canonical mass dimension of the operator and $\gamma_n$ its anomalous scaling, respectively. Singling out one specific reference operator (characterized by one fixed value $n$), this relation can be solved for $k$. Substituting the result into the scaling equation for the other operators leads to new scaling exponents defined with respect to the scaling of the reference operator. For example, selecting the reference operator to be the volume operator $\cO_0$ gives
\be\label{scalingexp}
\cO_n = \left(\cO_0 \right)^\frac{d_n + \gamma_n}{d_0 + \gamma_0} \, . 
\ee
Clearly, the $\gamma_n$ determine the new scaling exponents. This turns out to be useful when comparing to lattice simulations where the scaling ratio of two operators $\cO_0$ and $\cO_n$ may be accessed more easily than the coarse graining scale $k$.

Secondly, the recent proposal \cite{Klitgaard:2017ebu,Klitgaard:2018snm} for measuring the Ricci curvature on piecewise linear geometries resulting from the Causal Dynamical Triangulation program may allow to determine the value of the scaling exponents \eqref{scalingexp} from Monte Carlo simulations.

Thirdly, the scaling of operators within the class \eqref{C} has been investigated in detail by solving the Wetterich equation \cite{Wetterich:1992yh,Morris:1993qb,Reuter:1993kw,Reuter:1996cp} for gravitational actions of the form $\Gamma_k^{\rm grav} = \int d^4x \sqrt{g} f_k(R)$ with $f_k(R)$ an arbitrary, scale-dependent function of the Ricci scalar \cite{Codello:2007bd,Machado:2007ea,Benedetti:2012dx,Demmel:2012ub,Demmel:2013myx,Demmel:2014sga,Demmel:2014hla,Demmel:2015oqa,Dietz:2012ic,Dietz:2013sba,Dietz:2016gzg,Dietz:2015owa,Ohta:2015efa,Ohta:2015fcu,Alkofer:2018fxj,deBrito:2018jxt,Ohta:2018sze,Falls:2018ylp,Alkofer:2018baq,Burger:2019upn}. Thus the class \eqref{C} allows a direct comparison of the scaling properties found from analyzing the flow equation projected to $f(R)$-gravity and the composite operator formalism based on the Einstein-Hilbert action.

The remaining work is organized as follows. Sect. \ref{sect.2} provides a concise review of the background material, introducing the Wetterich equation, its solution based on the Einstein-Hilbert truncation, and the composite operator formalism. The computation of the anomalous dimensions for the geometric operators \eqref{C} is carried out in Sect. \ref{sect.3} and we discuss our findings in Sect.\ \ref{sect.4}. Technical details on the computation of the operator traces have been relegated to \ref{App.A} while some useful commutator identities on a background $d$-sphere are collected in
\ref{App.B}.
%------------------------------------------------------------------------
\section{Renormalization group flows and composite operators}
\label{sect.2}
%------------------------------------------------------------------------
We start by reviewing the Wetterich equation, its approximate solution given by the Einstein-Hilbert truncation and the composite operator formalism. Our discussion follows \cite{Reuter:2019byg} for which we also refer for more details.
%------------------------------------------------------------------------
\subsection{Wetterich equation and the Einstein-Hilbert truncation}
\label{sect.21}
%------------------------------------------------------------------------
An important tool for studying Asymptotic Safety is the Wetterich equation  for the effective average action $\Gamma_k$ \cite{Wetterich:1992yh,Morris:1993qb,Reuter:1993kw,Reuter:1996cp}
\be\label{FRGE}
\p_t \Gamma_k = \frac{1}{2} {\rm Tr}\left[ \left(\Gamma_k^{(2)} + \cR_k \right)^{-1}  \, \p_t \cR_k \right] \, , \quad \p_t \equiv k \p_k.
\ee
Here $k$ denotes the coarse graining scale, $\Gamma_k^{(2)}$  is the second functional derivative of $\Gamma_k$ with respect to the fluctuation fields, and ${\rm Tr}$ comprises a sum over fluctuation fields and an integral over loop-momenta. The infrared regulator $\cR_k$ provides a  $k$-dependent mass term for fluctuations with momentum $p^2 \lesssim k^2$ and vanishes for $p^2 \gg k^2$. As a consequence, the trace argument is peaked at momenta $p^2 \approx k^2$. Lowering $k$  unsuppresses further fluctuation modes, which are then integrated out. As a consequence the functional renormalization group equation (FRGE) \eqref{FRGE} implements Wilson's idea of renormalization by capturing the change of $\Gamma_k$ when an infinitesimal shell of fluctuation modes with momentum $p^2 \approx k^2$ is integrated out. Its solutions interpolate between the bare action for $k \rightarrow \infty$ and the effective action $\Gamma \equiv \Gamma_{k=0}$, provided that the corresponding limits exist.

An important property of the FRGE is that it permits constructing approximate solutions without having to rely on a small expansion parameter. The key idea is to formally expand $\Gamma_k$ in terms of interaction monomials multiplied by scale-dependent couplings $\bar{u}_i(k)$,
\be\label{actionfct}
\Gamma_k = \sum_{i=1}^{\infty} \, \bar{u}_i(k) \, \cO_i \, . 
\ee
Restricting the sum to a finite set, $i=1,\ldots,N$, substituting the resulting truncation of  $\Gamma_k$ into the FRGE, and projecting the resulting equation onto the interaction monomials  contained in the ansatz gives rise to a set of equations $k \p_k \bar{u}_i(k) = \bar{\beta}_{\bar{u}_i}(\{\bar{u}_j\}, k)$. Introducing dimensionless couplings $u_i(k) \equiv k^{-d_i} \bar{u}_i(k)$, where $d_i$ denotes the mass-dimension of the dimensionful coupling $\bar{u}_i(k)$, the flow can be recast in terms of the autonomous beta functions 
\be\label{beta_def}
\p_t u_i = \beta_{u_i}(\{u_j\}) \, , \qquad i = 1,\cdots, N \, . 
\ee

By definition, a fixed point $\{u_i^*\}$ is a point where all beta functions vanish simultaneously,
\be
\beta_{u_i}(\{u_j^*\}) = 0 \, . 
\ee
The flow in the vicinity of such a fixed point is conveniently studied by linearizing the system \eqref{beta_def}
\be\label{stabmat}
\p_t u_i \approx \sum_{j} \, {\bf B}_{i}{}^{j} (u_j - u_j^*) \, , \qquad {\bf B}_{i}{}^{j} \equiv \left. \frac{\p \beta_{u_i}}{\p u_j} \right|_{u_i = u_i^*} \, .
\ee
The eigenvalues $\lambda_i$ of the stability matrix ${\bf B}_{i}{}^{j}$ determine whether the renormalization group flow approaches (${\rm Re}\lambda_i < 0$) or is repelled from (${\rm Re}\lambda_i > 0$) the fixed point along the corresponding eigendirection as $k \rightarrow \infty$. Depending on whether the eigenvalues agree with the canonical dimension or receive quantum corrections one distinguishes between a Gaussian fixed point (GFP) or a non-Gaussian fixed point (NFGP).

Notably, the eigenvalues of the stability matrix ${\bf B}_{i}{}^{j}$ are invariant when redefining the couplings (coordinates) parameterizing the action functionals \eqref{actionfct}. This can be seen as follows. Consider a redefinition of the dimensionless couplings $u^\prime_i = u^\prime_i(u_j)$. Under this change the beta functions transform as a vector field
\be
\beta^\prime_{u^\prime_i}(u^\prime) = \frac{\p u^\prime_i}{\p u_j} \, \beta_{u_j}(u) \, . 
\ee
At a generic point $u$, the partial derivative ${\bf B}_{i}{}^{j}(u) = \frac{\p}{\p u_j} \beta_{u_i}$ then transforms according to
\be
{\bf B}^\prime_{i}{}^{j}(u^\prime) = \frac{\p u^\prime_i}{\p u_k} \, {\bf B}_k{}^l(u) \, \frac{\p u_l}{\p u^\prime_j}+ \beta_{u_l}(u) \, \frac{\p u_l}{\p u^\prime_j} \, \frac{\p^2 u^\prime_i}{\p u_k \p u_l} \, . 
\ee
At the fixed point $\beta_{u_i}(u_*) = 0$ and the second term on the right-hand side vanishes. As a consequence ${\bf B}^\prime_{i}{}^{j}(u^\prime_*)$ and ${\bf B}_{i}{}^{j}(u_*)$ are related by a similarity transformation, ensuring that their eigenvalues agree. Clearly, this property only holds for the eigenvalues and does not apply to the entries of the stability matrix.

%-----------------------------------------------------
\subsection{The Einstein-Hilbert truncation}
%-----------------------------------------------------
The arguably simplest projection of the Wetterich equation \eqref{FRGE} in the context of gravity approximates $\Gamma_k$ by the Einstein-Hilbert action supplemented by suitable gauge-fixing and ghost terms
\be\label{EHtrunc}
\Gamma_k \simeq \frac{1}{16 \pi G_k} \int d^dx \sqrt{g} \left[-R + 2 \Lambda_k \right] + \Gamma_k^{\rm gf} + S^{\rm ghost} \, . 
\ee
This ansatz retains two scale-dependent couplings, Newton's coupling $G_k$ and the cosmological constant $\Lambda_k$ and has been studied in detail \cite{Reuter:1996cp,Souma:1999at,Reuter:2001ag,Lauscher:2001ya,Litim:2003vp}. Following \cite{Reuter:1996cp}, the beta functions resulting from this ansatz are found as follows. First, we apply a linear split of the spacetime metric $g_{\mu\nu}$ into a background metric $\gb_{\mu\nu}$ and fluctuations $h_{\mu\nu}$
\be
g_{\mu\nu} = \gb_{\mu\nu} + h_{\mu\nu} \, . 
\ee
In order to simplify the computation, we choose the background metric to be the metric on the $d$-dimensional sphere, entailing that
\be\label{background}
\Rb_{\mu\nu\rho\sigma} = \frac{\Rb}{d(d-1)} \left[\gb_{\mu\rho} \gb_{\nu\sigma} - \gb_{\mu\sigma} \gb_{\nu\rho} \right] \, , \quad 
\Rb_{\mu\nu} = \frac{\Rb}{d} \, \gb_{\mu\nu} \, , \quad 
\Db_\mu \Rb = 0 \, . 
\ee
Covariant derivatives and curvature tensors constructed from $\gb_{\mu\nu}$ are denoted with a bar.
Subsequently, we chose harmonic gauge where
\be
\Gamma_k^{\rm gf} = \frac{1}{32\pi G_k} \int d^dx \sqrt{\gb} \gb^{\mu\nu} F_\mu F_\nu \, , \quad 
F_\mu = \Db^\nu h_{\mu\nu} - \tfrac{1}{2} \Db_\mu \gb^{\alpha\beta} h_{\alpha\beta} \, . 
\ee
Neglecting terms containing the fluctuation field $h_{\mu\nu}$, which are not required for the present computation, the ghost action for this gauge condition reads
\be
S^{\rm ghost} =  \sqrt{2} \int d^dx \sqrt{\gb} \, \bar{C}_\mu \, \left[ \Delta - \frac{1}{d} \Rb \right] C^\mu
\ee
where $\Delta \equiv - \gb^{\mu\nu} \Db_\mu \Db_\nu$. By decomposing the fluctuations into their trace and traceless part 
\be
h_{\mu\nu} = \widehat{h}_{\mu\nu} + \frac{1}{d} \, \gb_{\mu\nu} \, h \, , \qquad  \gb^{\mu\nu} \, \widehat{h}_{\mu\nu} = 0 \, , 
\ee
the propagators for the metric fluctuations can be cast into block-diagonal form
\be\label{EHprop}
\begin{split}
	\Gamma^{(2)}_k|_{\hh\hh} = & \frac{1}{32 \pi G_k} \left[\Delta - 2 \Lambda_k + C_T \Rb \right] \, ,  \\
	\Gamma^{(2)}_k|_{hh} = & - \frac{1}{32 \pi G_k} \, \frac{d-2}{2d} \, \left[\Delta - 2 \Lambda_k + C_S \Rb \right] \, ,
\end{split}
\ee
with
\be
C_T \equiv \frac{d(d-3)+4}{d(d-1)} \, , \qquad C_S \equiv \frac{d-4}{d} \, . 
\ee
The result for the ghost sector can be found in \cite{Reuter:1996cp,Reuter:2019byg}, but is not needed here. 

The regulator $\cR_k$ must provide a $k$-dependent mass term for fluctuations with (generalized) momentum smaller than $k^2$. We construct the matrix $\cR_k$ from the substitution rule for a Type I cutoff \cite{Codello:2008vh}, $\Delta \mapsto \Delta + R_k(\Delta)$. The resulting regulator 
is block-diagonal in field space. Its components in the gravitational sector are then given by
\be\label{EHreg}
\left. \cR_k \right|_{\hh\hh} = \frac{1}{32\pi G_k} R_k \, , \qquad \left. \cR_k \right|_{hh} = - \frac{1}{32 \pi G_k} \, \frac{d-2}{2d} \, R_k \, . 
\ee
In practice, we set $R_k = k^2 R^{(0)}(\Delta/k^2)$ with $R^{(0)}(z)$ being a dimensionless profile function interpolating between $\lim_{z\rightarrow\infty}R^{(0)}(z) = 0$ and $R^{(0)}(0) = 1$.
When presenting numerical results, we will work with the Litim regulator \cite{Litim:2000ci,Litim:2001up},
\be\label{Rlitim}
R^{(0)} = (1-z) \Theta(1-z) \, , 
\ee
where $\Theta(x)$ denotes the stepfunction.

The beta functions governing the change of $G_k$ and $\Lambda_k$ when successively integrating out fluctuations are found by substituting the ansatz \eqref{EHtrunc} into the Wetterich equation \eqref{FRGE} and projecting the flow onto the interaction monomials $\int d^dx\sqrt{\gb}$ and $\int d^dx \sqrt{\gb} \Rb$. This computation can be carried out rather straightforwardly by applying the early-time expansion of the heat-kernel along the lines \cite{Codello:2008vh}. The result is conveniently expressed in terms of the dimensionless couplings
\be
g_k = k^{d-2} \, G_k \, , \qquad \lambda_k = \Lambda_k \, k^{-2} \, ,
\ee
and reads
\be\label{betafct}
\p_t g_k = \left[d-2+\eta_N \right] g_k \, , \quad \p_t \lambda_k = \beta_\lambda(g_k,\lambda_k) \, . 
\ee
The explicit expressions for $\beta_\lambda$ and the anomalous dimension of the Newton coupling $\eta_N \equiv (G_k)^{-1} \p_t G_k$  have been computed in \cite{Reuter:1996cp} and read
\be
\begin{split}
	\beta_\lambda =  (\eta_N-2) \lambda  + \frac{g}{2(4\pi)^{d/2-1}} \Big[ &
	2d(d+1) \Phi^1_{d/2}(-2\lambda) - 8 d \Phi^1_{d/2}(0) % \\ & \; 
	- d(d+1)\eta_N \widetilde{\Phi}^1_{d/2}(-2\lambda)
	\Big] \, ,
\end{split}
\ee
and
\be
\eta_N(g,\lambda) = \frac{g B_1(\lambda)}{1 - g B_2(\lambda)}
\ee
with
\be
\begin{split}
	B_1(\lambda) = & \, \frac{1}{3(4\pi)^{d/2-1}} \Big[d(d+1) \Phi^1_{d/2-1}(-2\lambda) - 6d(d-1) \Phi^2_{d/2}(-2\lambda) \\ 
	& \qquad \qquad \qquad - 4 d \Phi^1_{d/2-1}(0) - 24 \Phi^2_{d/2}(0) \Big] \, ,  \\
	B_2(\lambda) = & \, -\frac{1}{6(4\pi)^{d/2-1}} \left[ d(d+1) \widetilde{\Phi}^1_{d/2-1}(-2\lambda) - 6d(d-1) \widetilde{\Phi}^2_{d/2}(-2\lambda)  \right] \, . 
\end{split}
\ee
The dimensionless threshold functions $\Phi^p_n(w)$ and $\widetilde{\Phi}^p_n(w)$ are defined as
\be\label{threshold}
\begin{split}
	\Phi^p_n(w) \equiv & \, \frac{1}{\Gamma(n)} \int_0^\infty dz \, z^{n-1} \, \frac{R^{(0)}(z) - z R^{(0)\prime}(z)}{[z + R^{(0)}(z) + w]^p} \, , \\
	\widetilde{\Phi}^p_n(w) \equiv & \, \frac{1}{\Gamma(n)} \int_0^\infty dz \, z^{n-1} \, \frac{R^{(0)}(z) - z R^{(0)\prime}(z)}{[z + R^{(0)}(z) + w]^p} \, , 
\end{split}
\ee
and encode the regulator-dependence of the flow. For the regulator \eqref{Rlitim} the integrals  can be performed analytically, yielding
\be\label{thresholdLitim}
\Phi^{p, {\rm Litim}}_n(w) = \frac{1}{\Gamma(n+1)} \, \frac{1}{(1+w)^p} \, , \quad
\widetilde{\Phi}^{p, {\rm Litim}}_n(w) = \frac{1}{\Gamma(n+2)} \, \frac{1}{(1+w)^p}\, . 
\ee
For later convenience we also introduce the combination
\be\label{qfcts}
q^p_n(w) \equiv \Phi^p_n(w) - \frac{1}{2} \, \eta_N \, \widetilde{\Phi}^p_n(w) \, . 
\ee

The system \eqref{betafct} possesses a GFP situated at the origin $\{g_*,\lambda_*\} = 0$. In addition one also finds a NGFP. For the regulator \eqref{Rlitim} the later is situated at
\be\label{NGFPpos}
\begin{split}
	d=3: & \quad \{g_*,\lambda_*\} = \{\, 0.199 \, , \, 0.063 \, \} \, , \\
	d=4: & \quad \{g_*,\lambda_*\} = \{\, 0.707 \, , \, 0.193 \,  \} \, . 
\end{split}
\ee
Evaluating the stability matrix \eqref{stabmat} at these NGFPs yields
\be\label{stabmatex}
\left. {\bf B}_{i}{}^{j} \right|_{d=3} =
\left[
\begin{array}{cc}
	-1.258	& 1.225 \\
	-0.569	& -1.037
\end{array}
\right] 
\, , \quad
\left. {\bf B}_{i}{}^{j} \right|_{d=4} =
\left[
\begin{array}{cc}
	-0.608	& 0.959 \\
	-10.44	& -2.342
\end{array}
\right] \, . 
\ee 
The ${\bf B}_{i}{}^{j}$ are neither symmetric nor ``almost diagonal'' in the sense that the off-diagonal elements are small compared to the diagonal ones. The computation of the eigenvalues is straightforward and yields 
\be\label{StabilityMat}
\begin{split}
	d=3: & \quad \lambda_{1,2} = -1.148 \pm 0.827 i \, ,  \\
	d=4: & \quad \lambda_{1,2} = -1.475 \pm 3.043 i \, .  
\end{split}
\ee
Thus, in both cases, the NGFP constitutes a spiraling UV-attractor which captures RG trajectories in its vicinity in the limit $k \rightarrow \infty$. The existence and qualitative properties of the NGFP are stable with respect to changes in the regulator \cite{Reuter:2001ag,Gies:2015tca}.

\begin{figure}[t]
	\centering
	\includegraphics[width=6.8cm]{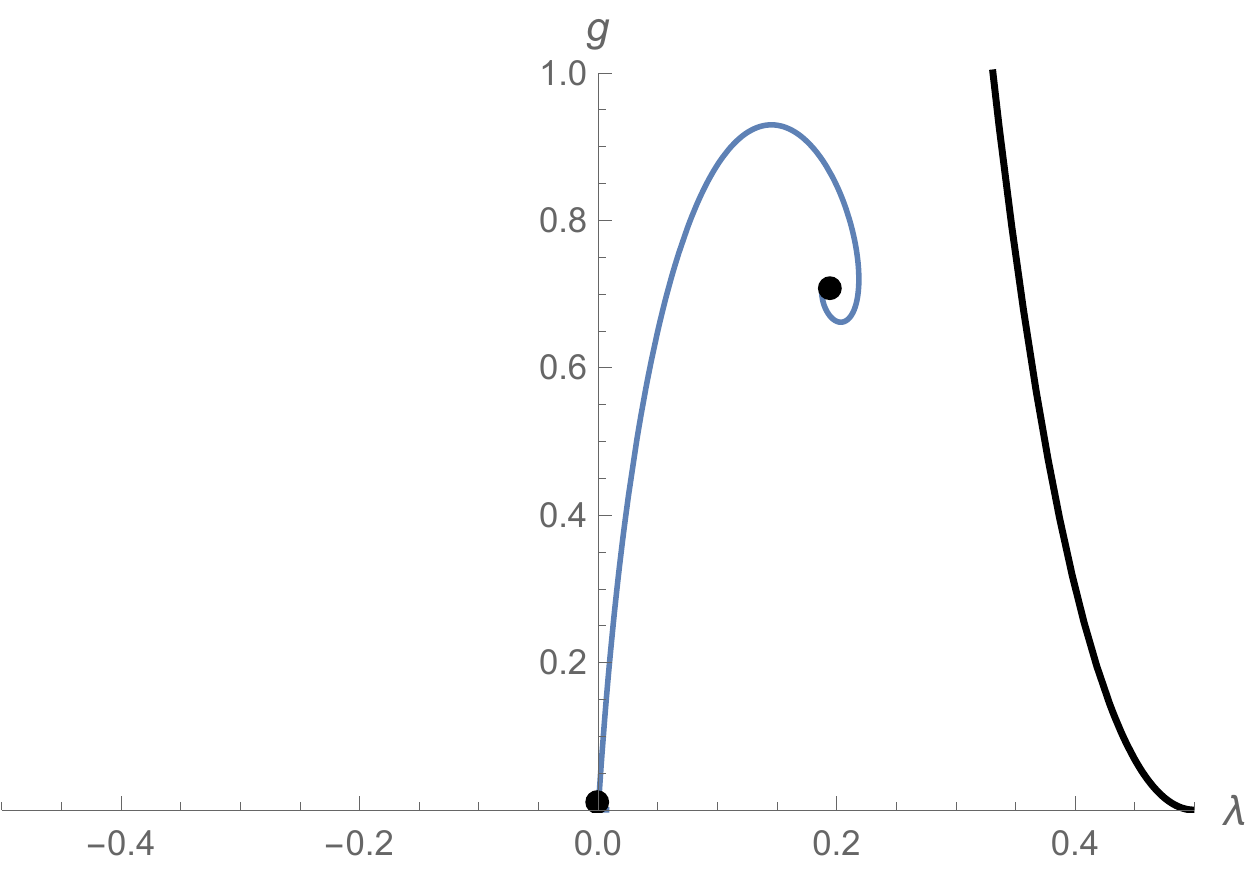} \\[2ex]
	\includegraphics[width=0.48\textwidth]{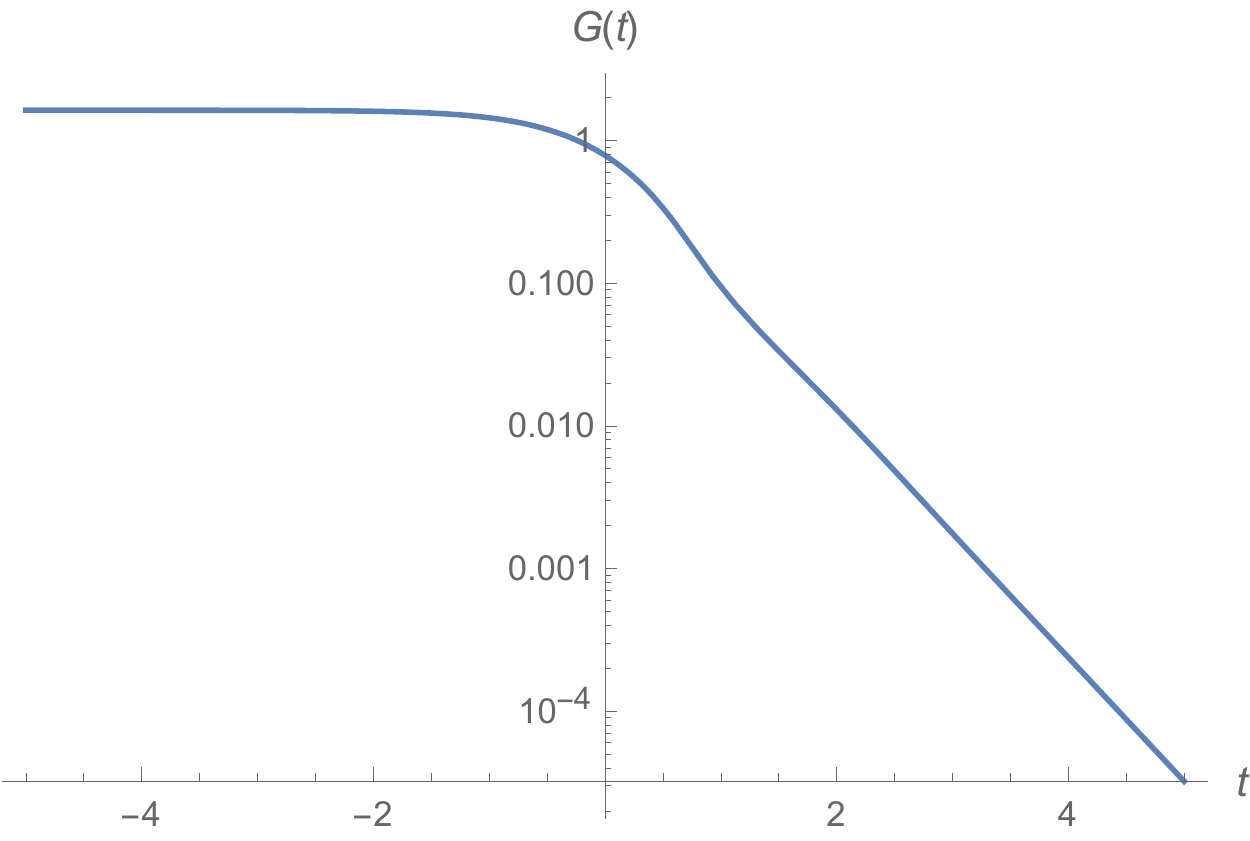}
	\includegraphics[width=0.48\textwidth]{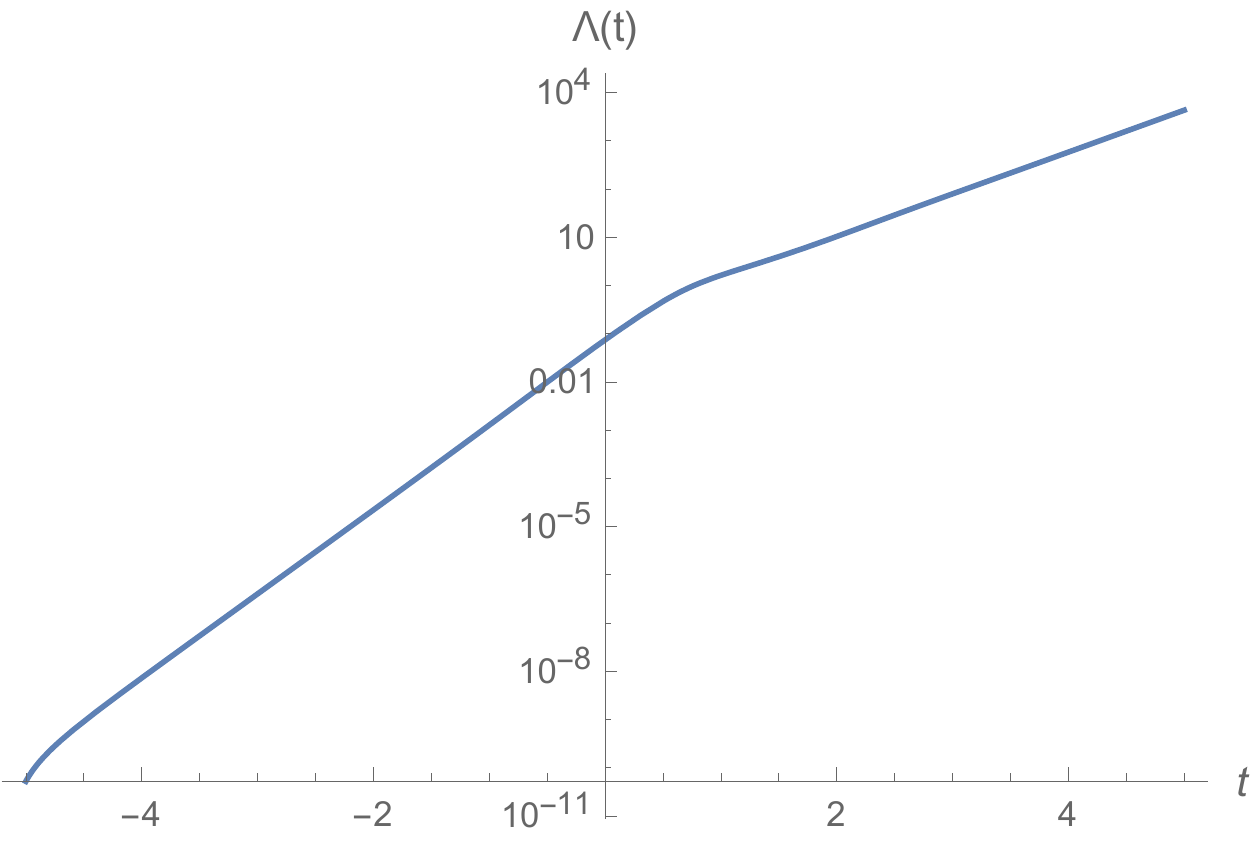}
	\caption{\label{Fig.1} Top: Solution of the system \eqref{betafct} for $d=4$ connecting the Gaussian and non-Gaussian fixed points marked as black dots. The black line emanating from the point $(\lambda,g) = (1/2,0)$ constitutes a singular locus of the beta functions where $\eta_N$ diverges. For a complete classification of the RG trajectories arising within the Einstein-Hilbert truncation see \cite{Reuter:2001ag}. Bottom: the dimensionful couplings $\Lambda_k$ and $G_k$ as a function of the RG time $t \equiv \ln(k/k_0)$. The solutions interpolate between the fixed point regime where $G_k = g_*/k^2$, $\Lambda_k = \lambda_* k^2$ and the classical regime where Newton's coupling freezes to a constant value. The scale $k_0$ is chosen such that the transition between the regimes is at $k \approx k_0$.}
\end{figure}

The system \eqref{betafct} is conveniently solved numerically. A solution of specific interest is the one which connects the NGFP at large values of $k$ and the GFP for $k \rightarrow 0$. For $d=4$ this solution is shown in Fig.\ \ref{Fig.1}. 
As its characteristic feature, the value of Newton's constant, $G_0 \equiv \lim_{k \rightarrow 0} G_k$, is a free constant while the comological constant $\Lambda_0 \equiv \lim_{k \rightarrow 0} \Lambda_k = 0$ vanishes. We will evaluate the anomalous dimensions $\gamma_n$ along this solution later on.
%-----------------------------------------------------
\subsection{Composite operator formalism}
%-----------------------------------------------------
The key virtue of the composite operator formalism is that it allows computing the anomalous dimension of generic operators $\cO$ which are not part of the ansatz made for the effective average action $\Gamma_k$. Following the exposition \cite{Pagani:2016dof}, we now explain the main steps leading to the master equation \cite{Pagani:2016dof,Pawlowski:2005xe}
\be\label{CompositeMaster}
\gamma_n \, \cO_n = - \frac{1}{2} \left. {\rm Tr} \left[ \left(\Gamma_k^{(2)} + \cR_k \right)^{-1} \, \cO^{(2)}_n \, \left(\Gamma_k^{(2)} + \cR_k \right)^{-1} \, \p_t \cR_k       \right] \right|_{\cO} \, . 
\ee

As starting point, consider the expectation value of an operator $\cO$, constructed from a suitably gauge-fixed and regulated partition function. Symbolically,
\be
\begin{split}
	\langle \cO(x) \rangle \equiv & \, \cN \int \cD \chi \, \cO(x) \, e^{-S[\chi]} \\
	= & - \frac{\delta}{\delta \epsilon(x)} \, \cN \int \cD \chi \, e^{-S[\chi] - \epsilon \cdot \cO} \, \big|_{\epsilon = 0} \, . 
\end{split} 
\ee
Here $\cN$ is a normalization constant and we introduced $\epsilon(x)$ as the source associated with the new operator. The generating functional for the connected Green functions $W[J;\epsilon]$ is  defined via
\be
e^{W[J;\epsilon]} = \int \cD \chi \, e^{-S[\chi] + J \cdot \chi - \epsilon \cdot \cO} \, . 
\ee
The effective action $\Gamma[\varphi; \epsilon]$ depends on the classical fields $\varphi \equiv \frac{\delta W}{\delta J}$ and is given by the Legendre transform of $W[J;\epsilon]$ 
\be
\Gamma[\varphi; \epsilon] = J \cdot \varphi - W[J;\epsilon] \, . 
\ee
Since $J$ and $\varphi$ are independent of the source $\epsilon$, this definition entails
\be
\frac{\delta \Gamma[\varphi; \epsilon]}{\delta \epsilon} = - \frac{\delta W[J; \epsilon]}{\delta \epsilon} \, . 
\ee

The generalization of this construction to the effective average action is then straightforward. Including the new source $\epsilon$ in $\Gamma_k[\varphi; \epsilon]$ and taking the variation of the Wetterich equation \eqref{FRGE} with respect to the source yields \cite{Ellwanger:1994iz,DAttanasio:1996tzp,Litim:1998qi,Pagani:2016dof}
\be\label{opint}
\p_t \left. \big( \frac{\delta}{\delta \epsilon} \Gamma_k[\varphi; \epsilon] \big) \right|_{\epsilon = 0} = 
- \frac{1}{2} \left. {\rm Tr} 
\left[\big( \Gamma_k^{(2)} + \cR_k \big)^{-1} \, \frac{\delta \Gamma^{(2)}_k}{\delta \epsilon} \, \big( \Gamma_k^{(2)} + \cR_k \big)^{-1} \p_t \cR_k
\right]
\right|_{\epsilon = 0} . 
\ee
Thus the scale-dependence of the operator $\cO$ is determined by a single operator insertion of $\cO^{(2)}$ into the loop building the right-hand side of the Wetterich equation.

We now recast \eqref{opint} into a more practical form, by first introducing a set of $k$-dependent operators $[\cO_k]_i$, $i=1,\ldots,N$, which we relate to the effective average action by
\be\label{eq36}
[\cO_k]_i = \frac{\delta}{\delta \epsilon_i} \Gamma_k[\varphi; \epsilon_1, \cdots, \epsilon_N] |_{\epsilon = 0}\, . 
\ee
It is then convenient to express the scale-dependent $[\cO_k]_i$ in terms of a $k$-independent basis
\be\label{eq37}
[\cO_k]_i = \sum_{j=1}^N \, Z_{ij}(k) \, \cO_j \, . 
\ee
Here the matrix $Z_{ij}(k)$ is reminiscent of a wave-function renormalization associated with the operators $\cO_j$. This suggests to define the (matrix-valued) anomalous dimension of the set of operators according to
\be\label{eq38}
\gamma_{ij} \equiv \left( Z^{-1} \p_t Z \right)_{ij} \, . 
\ee
Substituting the relation \eqref{eq36} together with the definitions given in eqs.\ \eqref{eq37} and \eqref{eq38} into \eqref{opint} then yields the following equation for the anomalous dimensions $\gamma_{ij}$:
\be\label{CompositeMaster2}
\sum_{j=1}^N \gamma_{ij} \, \cO_j = - \frac{1}{2} \left. {\rm Tr} \left[ \left(\Gamma_k^{(2)} + \cR_k \right)^{-1} \, \cO^{(2)}_i \, \left(\Gamma_k^{(2)} + \cR_k \right)^{-1} \, \p_t \cR_k       \right] \right|_{\cO} \, . 
\ee
Restricting the set $\{\cO_i\}$ to a single operator $\cO$, this relation simplifies to the master equation \eqref{CompositeMaster}. In the next section, we use this equation as the starting point for computing the anomalous dimension of the geometric operators \eqref{C}.

%-----------------------------------------------------
\section{The anomalous scaling dimension of geometric operators}
\label{sect.3}
%-----------------------------------------------------
In this section we utilize the composite operator equation \eqref{CompositeMaster} to compute the anomalous dimensions $\gamma_n$ associated with the geometric operators \eqref{C}. Our main result is derived in subsection \ref{sect.3.1}, the resulting characterization of the NGFP and the emergence of  classical geometry are discussed in subsection \ref{sect.3.2}, and we conclude with a comparison to results obtained from $f(R)$-type truncations in subsection \ref{sect.3.4}.
%-----------------------------------------------------
\subsection{Computing the anomalous dimension $\gamma_n$}
\label{sect.3.1}
%-----------------------------------------------------
The goal of this subsection is to evaluate \eqref{CompositeMaster} for the class of geometric operators given in eq.\ \eqref{C}. The propagators are approximated by the Einstein-Hilbert action and given in eq.\ \eqref{EHprop}.

We start by expanding $\cO_n$ in terms of fluctuations in the background \eqref{background}. The term quadratic in $h_{\mu\nu}$ is
\be\label{d2ops}
\begin{split}
	\delta^2\cO_n = & \,  \int d^dx \sqrt{\gb}  \Rb^{n-2} \Big[
	\hh_{\mu\nu} \left[ - \tfrac{n}{2} \Delta \Rb + \left( n \tfrac{d-2}{d(d-1)} - \tfrac{1}{2} \right) \Rb^2 \right] \hh^{\mu\nu} \\
	& - \hh_{\mu\nu} \left[n \, \Rb \, \delta^\mu_\alpha \Db^\nu \Db_\beta - n(n-1) \Db^\mu \Db^\nu \Db_\alpha \Db_\beta \right] \hh^{\alpha\beta} \\
	& + h\left[ n(n-1) \tfrac{(d-1)^2}{d^2} \Delta^2 + n \tfrac{d^2 - (4n-1)d +4n-2}{2d^2} \Rb \Delta + \left( \tfrac{d-2}{4d} - \tfrac{n(d-n-1)}{d^2} \right) \Rb^2  \right] h \\
	&
	+ h \left[ \left( n \tfrac{d-2n}{d} \Rb + 2 n (n-1) \tfrac{d-1}{d} \Delta \right) \Db_\mu \Db_\nu \right] \hh^{\mu\nu}
	\Big] \, . 
\end{split}
\ee
Since the gravitational propagators and regulators are diagonal once $h_{\mu\nu}$ is decomposed into its traceless and trace-part, the off-diagonal term containing $h$ and $\hh_{\mu\nu}$ does not contribute in the computation of the anomalous dimension. Thus the two relevant matrix elements of $\cO_n^{(2)}$ are
\ba\label{Ohh}
\left. \cO^{(2)}_n \right|_{hh}  & = &  c_0^S \, \Rb^n + c_1^S \, \Delta \, \Rb^{n-1} + c_2^S \, \Delta^2 \, \Rb^{n-2}  \, , \\ \label{Ohtht}
\left[\left. \cO^{(2)}_n \right|_{\hh\hh}\right]^{\mu\nu}{}_{\alpha\beta}  & = & 
\big[ c_0^T \, \Rb^n + c_1^T \, \Delta \, \Rb^{n-1}  \big] \big[ {\mathbbm 1}_T \big]^{\mu\nu}{}_{\alpha\beta} + \left[\left. \cO^{(2),{\rm nm}}_n \right|_{\hh\hh}\right]^{\mu\nu}{}_{\alpha\beta}
\ea
with the non-minimal (nm)-part in the tensor sector is given by
\be\label{Ohthtnonmin}
\left[\left. \cO^{(2),{\rm nm}}_n \right|_{\hh\hh}\right]^{\mu\nu}{}_{\alpha\beta} = 
- n \, \Rb^{n-1} \, \delta^{(\mu}_{(\alpha} \Db^{\nu)} \Db_{\beta)} + n(n-1) \Rb^{n-2} \, \Db^{(\mu} \Db^{\nu)} \Db_{(\alpha} \Db_{\beta)} \, .
\ee
Here 
\be\label{unitT}
\left[ {\mathbbm 1}_T \right]^{\mu\nu}{}_{\alpha\beta} \equiv \frac{1}{2} \left[ \delta^\mu_\alpha \delta^\nu_\beta - \delta^\mu_\beta \delta^\nu_\alpha \right] - \frac{1}{d} \, \gb^{\mu\nu} \, \gb_{\alpha\beta}
\ee
denotes the unit on the space of traceless symmetric matrices and symmetrization is with unit strength $(\alpha\beta) = (\alpha\beta + \beta\alpha)/2$. The coefficients $c_i^S(d,n)$ and $c_i^T(d,n)$ are
\be
\begin{split}
	& c_0^T = \, \frac{n(d-2)}{d(d-1)} - \frac{1}{2} \, , \qquad c_1^T = - \frac{n}{2} \, , \\
	& c_0^S = \, \frac{d-2}{4d} - \frac{n(d-n-1)}{d^2} \, , \quad
	c_1^S =  \, \frac{n}{2d^2} \left( d^2 - (4n-1)d +4n-2 \right) \, , \\
	& c_2^S =  \, \frac{(d-1)^2}{d^2} \, n(n-1) \, .
\end{split}
\ee

At this stage, all ingredients to construct \eqref{CompositeMaster} for the geometric operators \eqref{C} are available. The block-diagonal form of the regulator $\cR_k$, eq.\ \eqref{EHreg}, entails that the trace over fluctuation fields decomposes into a sum of the scalar and tensor fluctuations
\be\label{mastertraces}
\gamma_n \, \cO_n = - \frac{1}{2} \left( \left. 
{\rm Tr}_T \left[  \cO^{(2)}_n \big|_{\hh\hh} \, W_T(\Delta)     \right] +
{\rm Tr}_S \left[  \cO^{(2)}_n \big|_{hh} \, W_S(\Delta)     \right] \right) \right|_{\cO_n} \, .
\ee
The operator-valued functions $W^{T}(\Delta)$ and $W^S(\Delta)$ are functions of the Laplacian and contain the propagators and regulators of the corresponding sectors
\be\label{WTdef}
\begin{split}
	W_T(\Delta) \equiv & \, \left[  \Gamma^{(2)}_k \big|_{\hh\hh} +  \cR_k \big|_{\hh\hh} \right]^{-2} \, \p_t  \cR_k \big|_{\hh\hh} \, , \\ 
	W_S(\Delta) \equiv & \, \left[  \Gamma^{(2)}_k\big|_{hh} +  \cR_k \big|_{hh} \right]^{-2} \, \p_t  \cR_k \big|_{hh} \, .
\end{split}
\ee
Their explicit expressions are obtained by substituting the propagators \eqref{EHprop} and regulators \eqref{EHreg}:
\be\label{WTev}
\begin{split}
	W_T(\Delta) \equiv & \, 32 \pi G_k  \left[ \Delta + R_k - 2 \Lambda_k + C_T \Rb \right]^{-2} \left(\p_t R_k - \eta_N R_k \right) \, ,  \\
	W_S(\Delta) \equiv & \, - 32 \pi G_k \, \frac{2d}{d-2} \, \left[ \Delta + R_k - 2 \Lambda_k + C_S \Rb \right]^{-2} \left(\p_t R_k - \eta_N R_k \right) \, .
\end{split}
\ee
The traces \eqref{mastertraces} are then readily evaluated using the early-time heat-kernel in combination with Laplace-transform techniques. We provide the technical details of this computation in \ref{App.A}.

From the structure of the operator insertions, one infers that the anomalous dimensions $\gamma_n$ are second order polynomials in $n$
\be\label{mainresult}
\gamma_n = \frac{32 \pi g}{(4\pi)^{d/2}} \, \left(A_0(\lambda) + A_1(\lambda) \, n + A_2(\lambda) \, n^2 \right) \, .
\ee
The coefficients $A_i$ depend on the dimensionless couplings $g,\lambda$ as well as the dimension of spacetime $d$. They are conveniently given in terms of the threshold functions $q^p_n \equiv q^p_n(-2\lambda)$ defined in eq.\ \eqref{qfcts}:
\be
\begin{split}
	A_0(\lambda) = & \, \frac{1}{4} d(d+1) \, q^2_{d/2} \, , \\
	A_1(\lambda) = & \, d_1^0 \, q^2_{d/2} + d_1^1 \, q^3_{d/2+1} + d_1^2 \, q^4_{d/2+2} \, , \\
	A_2(\lambda) = & \, d_2^0 \, q^2_{d/2} + d_2^1 \, q^3_{d/2+1} + d_2^2 \, q^4_{d/2+2} \, .
\end{split}
\ee
The $d$-dependent coefficients entering $A_1(\lambda)$ and $A_2(\lambda)$ are
\be
\begin{split}
	d_1^0 = & \, \tfrac{
		30 d^6 - 185 d^5 -528 d^4 +2515 d^3 - 1500 d^2 -3500  d +2448 }{1440 d (d-1) (d-2)  } \, , \\
	d_1^1 = & - \tfrac{3 d^6 - 19 d^5 + 67 d^4 - 109 d^3 -94 d^2 + 392 d - 288}{12 d (d-1) (d-2)} \, , \\
	d_1^2 = & - \tfrac{3}{4} \, \tfrac{d^5  - 12 d^4 + 29 d^3 +26 d^2 - 128 d + 96}{d (d-1) (d-2) } \, , \\
	d_2^0 = & \, \tfrac{5 d^5 - 162 d^4 -355 d^3 +1620 d^2 +2060d  -2448}{1440 d (d-1) (d-2)} \, , \\
	d_2^1 = & \, -  \tfrac{ d^5 -22  d^4 + 31 d^3 + 190 d^2 -440 d +288  }{12 d (d-1) (d-2)} \, , 
\end{split}
\ee
together with $d_2^2 = - d_1^2$. This formula constitutes our main result. 
For $n=0$ it is in agreement with the earlier derivation of the anomalous scaling dimension of the spacetime volume \cite{Pagani:2016dof}.

%-----------------------------------------------------
\subsection{The anomalous dimension on the $g$-$\lambda$-plane}
\label{sect.3.2}
%-----------------------------------------------------
The central property of eq.\ \eqref{mainresult} is that it assigns values $\gamma_n$, $n=0,1,2,\cdots$, \emph{to every point on the $\lambda$-$g$--plane.} Remarkably, this assignment is independent of any specific solution of the flow equation \eqref{betafct}. The dependence on $\lambda$ and $g$ is inherited from the parametric form of the gravitational propagators without reference to a RG flow.

\begin{figure}[t]
	\centering
	\includegraphics[width=0.48\textwidth]{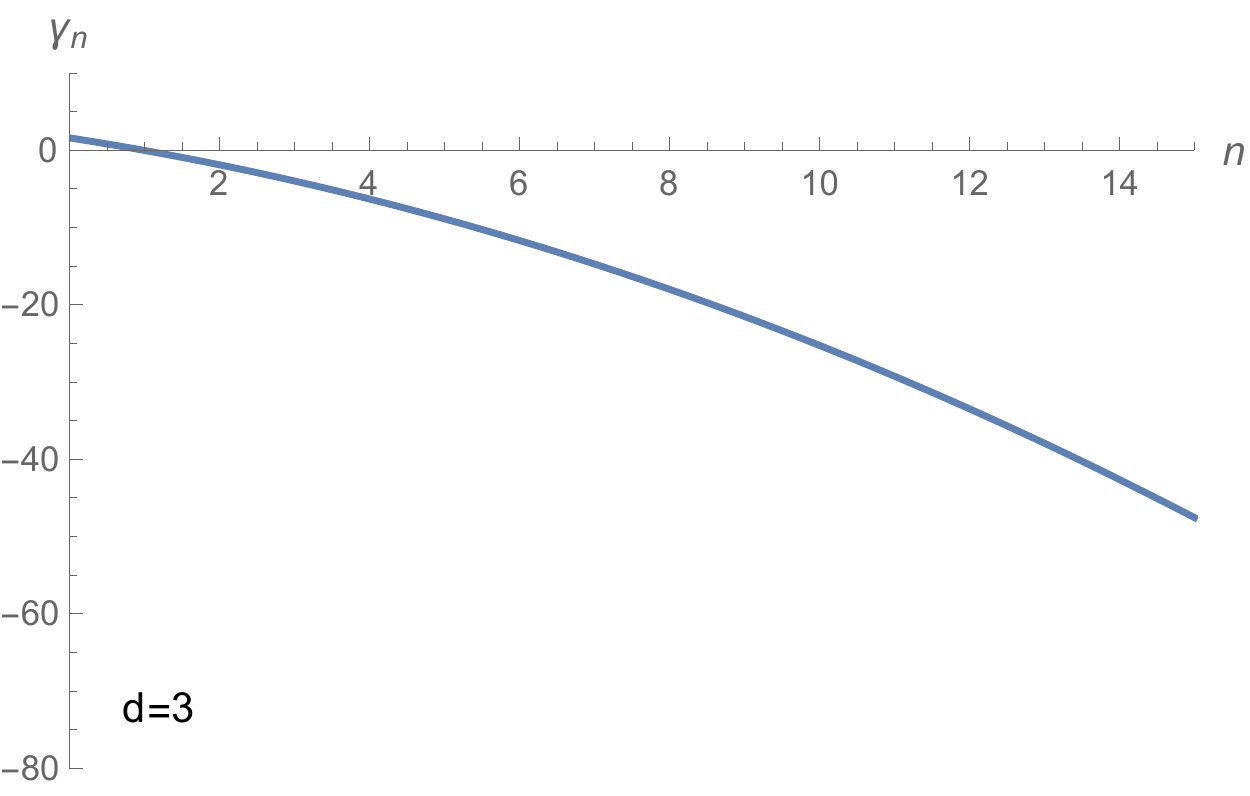}
	\includegraphics[width=0.48\textwidth]{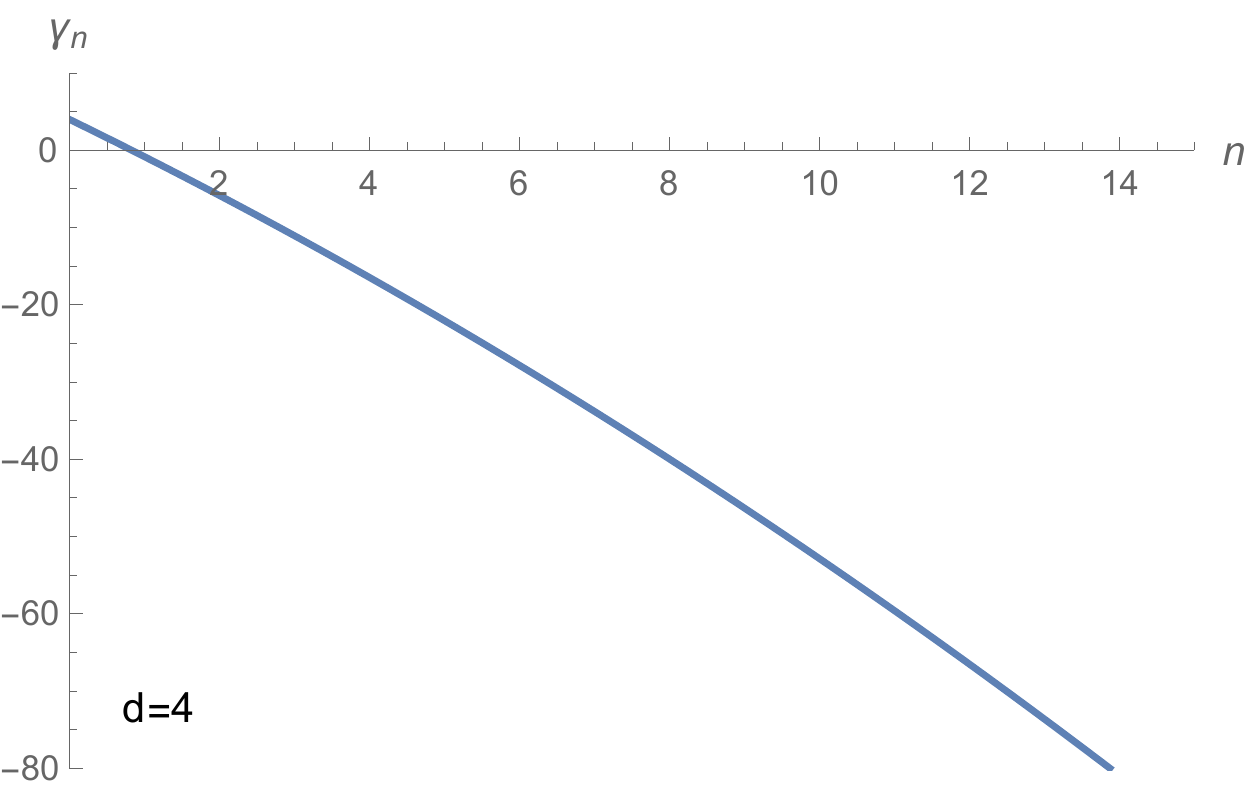}
	\caption{\label{Fig.3} Illustration of the anomalous dimensions \eqref{ScalingMaster} in $d= 3$ (left) and $d= 4$ (right).  The anomalous dimension decreases monotonically with increasing values $n$.}
\end{figure}

It is then straightforward to evaluate this map at specific points in the $\lambda$-$g$--plane. At the GFP, $(\lambda,g) = (0,0)$, the vanishing of $g$ entails that all anomalous dimensions are zero
\be
\mbox{GFP:} \qquad \gamma_n = 0 \, , \qquad \forall \, n,d \, . 
\ee
Thus, for $g=0$ one recovers a classical geometry where all geometric operators scale with their corresponding mass-dimension. At the fixed points \eqref{NGFPpos} one finds
\be\label{ScalingMaster}
\begin{array}{ll}
	d=3 \, : \qquad \gamma_n^* = 1.591 - 1.505 n - 0.118 n^2 \, , \\
	d=4 \, : \qquad \gamma_n^* = 3.987 - 4.733 n - 0.095 n^2 \, .
\end{array}
\ee
This set of characteristic numbers earmarks the quantum spacetime associated with the
non-Gaussian fixed points. From \eqref{mainresult}, it is clear that the $n$-dependence of the  anomalous dimensions is given by second-order polynomials in $n$. For $d=3$ and $d=4$, these polynomials are shown in Figure \ref{Fig.3}.
The $\gamma_n$ are positive for $n=0$ only. For $n \ge 1$ the anomalous dimension turns out to be negative. Surprisingly, the numerical coefficients multiplying the $n^2$-terms are comparatively small, so that $\gamma_n$ essentially follows a linear behavior. The linear and quadratic terms have equal values for $n \lesssim 49$ $(d=4)$ ($n \lesssim 12$ for $d=3$). This gives a good indication where deviations from the linear behavior set in. 

\begin{figure}[h!]
	\centering
	\includegraphics[width=0.47\textwidth]{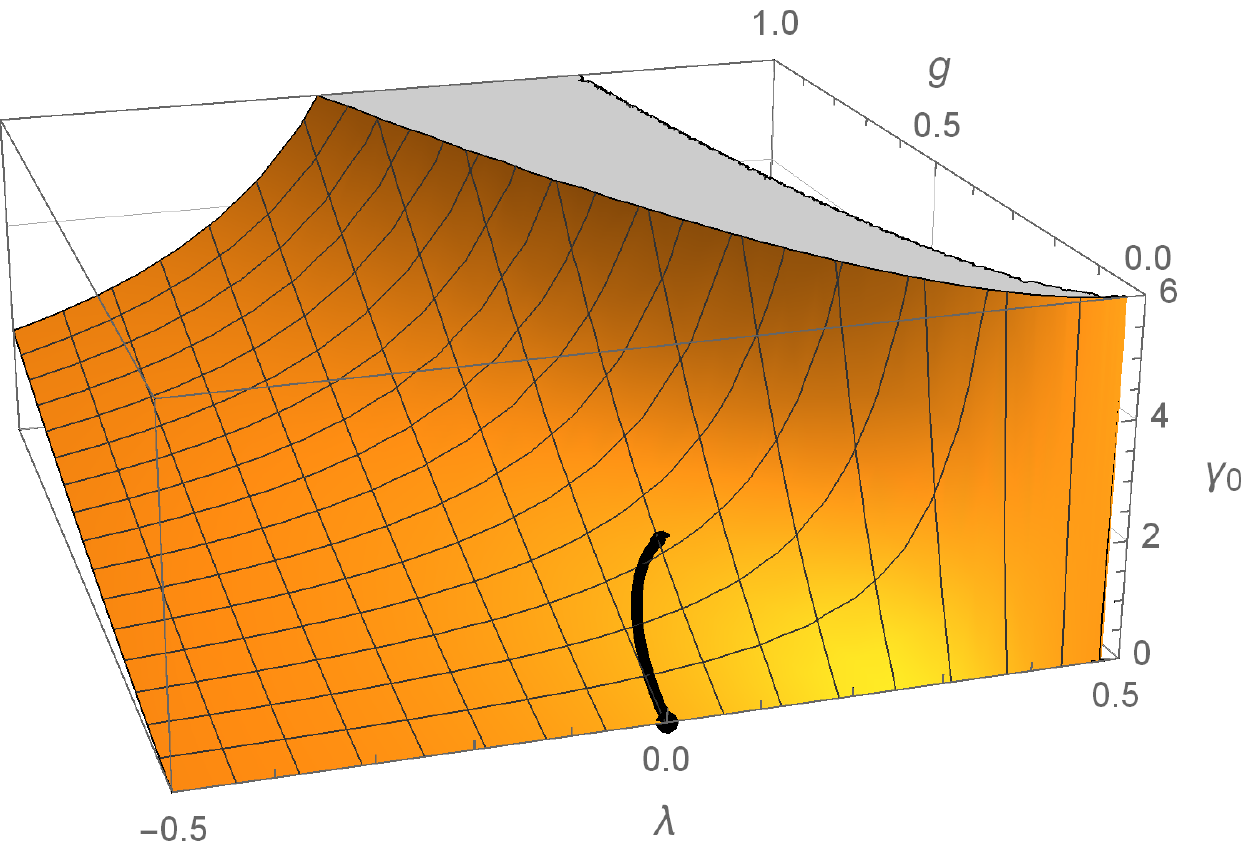} \;
	\includegraphics[width=0.47\textwidth]{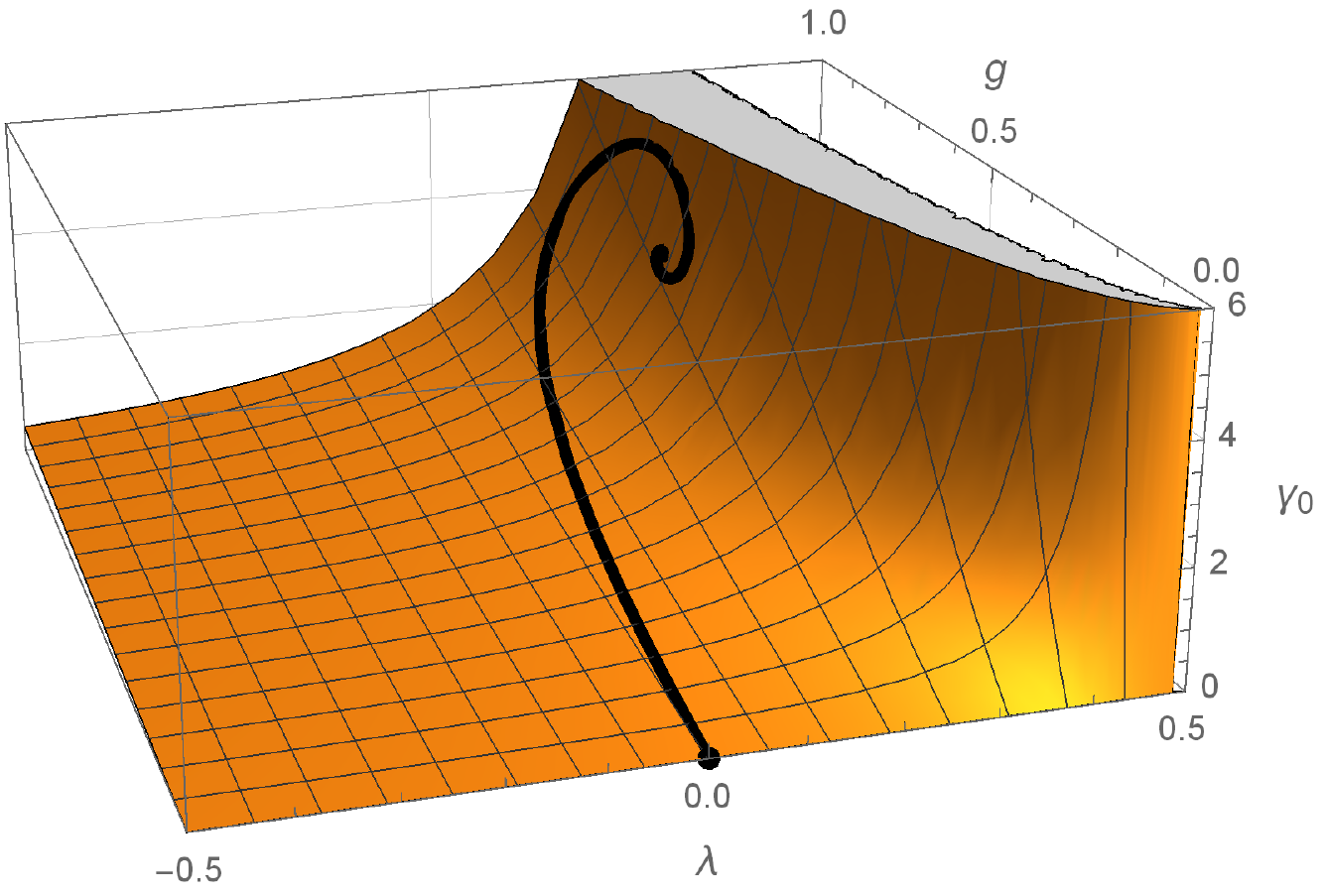} \\[1.2ex]
	\includegraphics[width=0.47\textwidth]{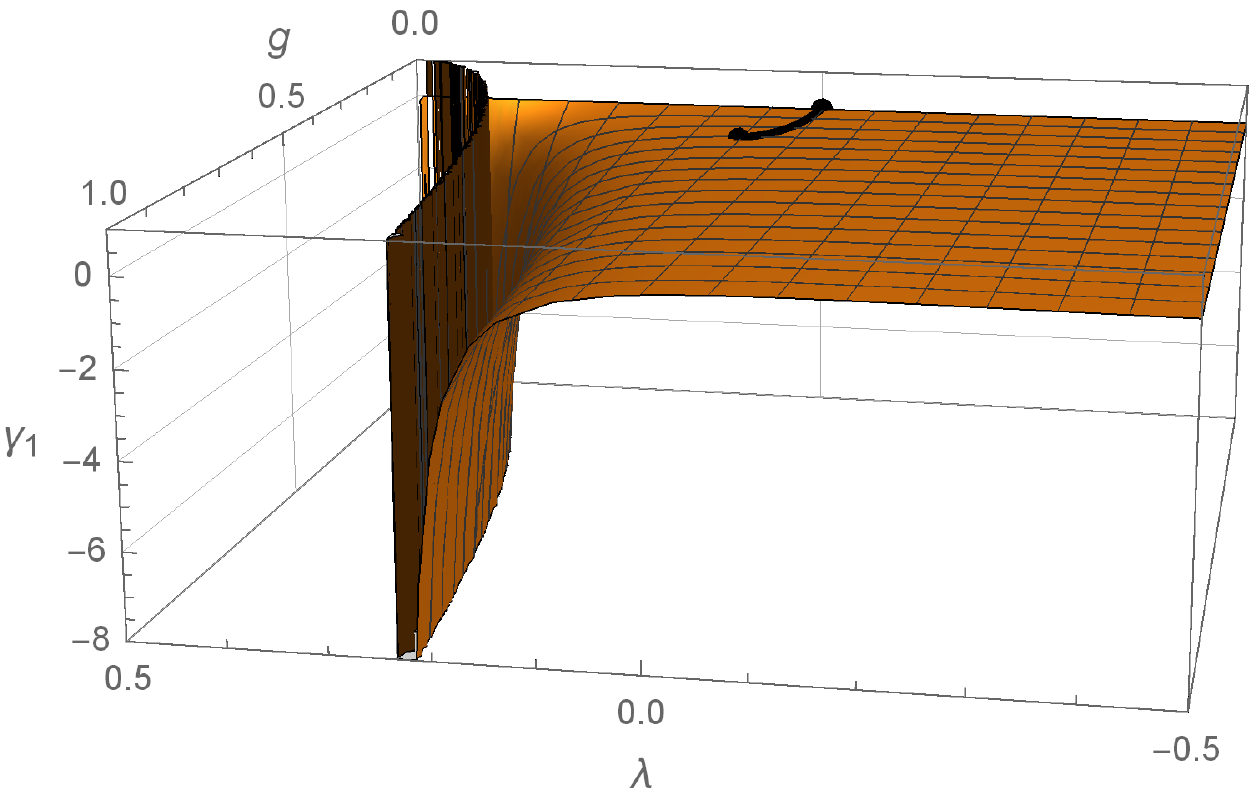} \;
	\includegraphics[width=0.47\textwidth]{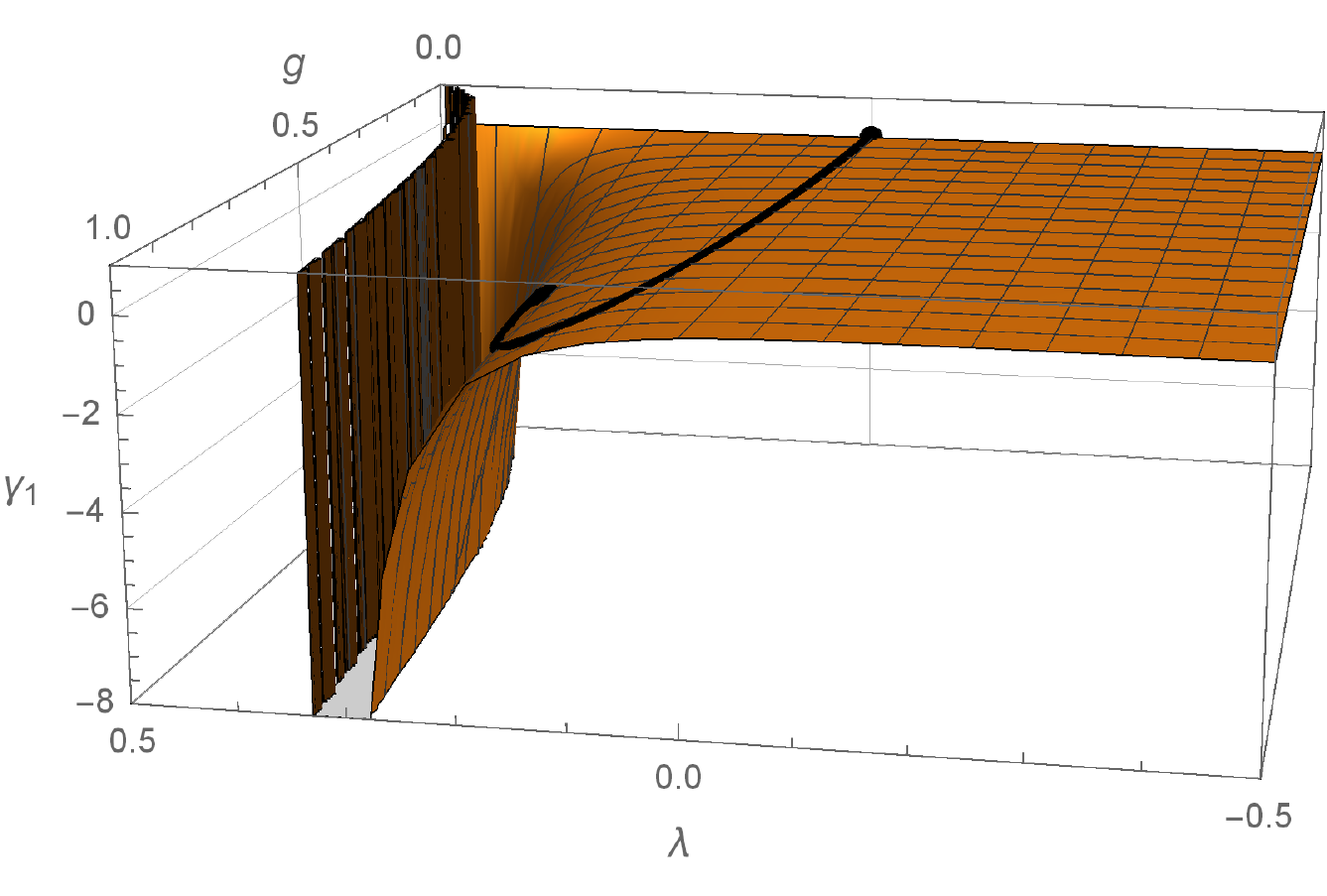} \\[1.2ex]
	\includegraphics[width=0.47\textwidth]{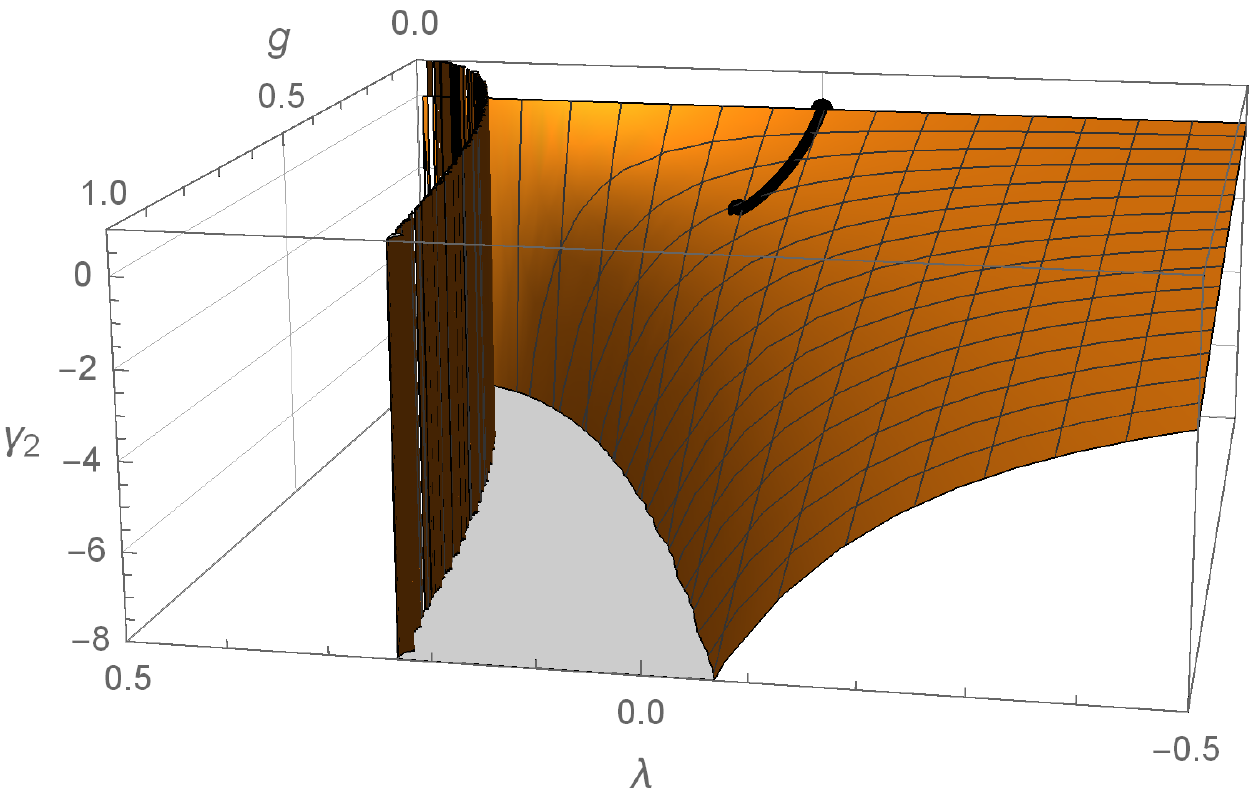} \;
	\includegraphics[width=0.47\textwidth]{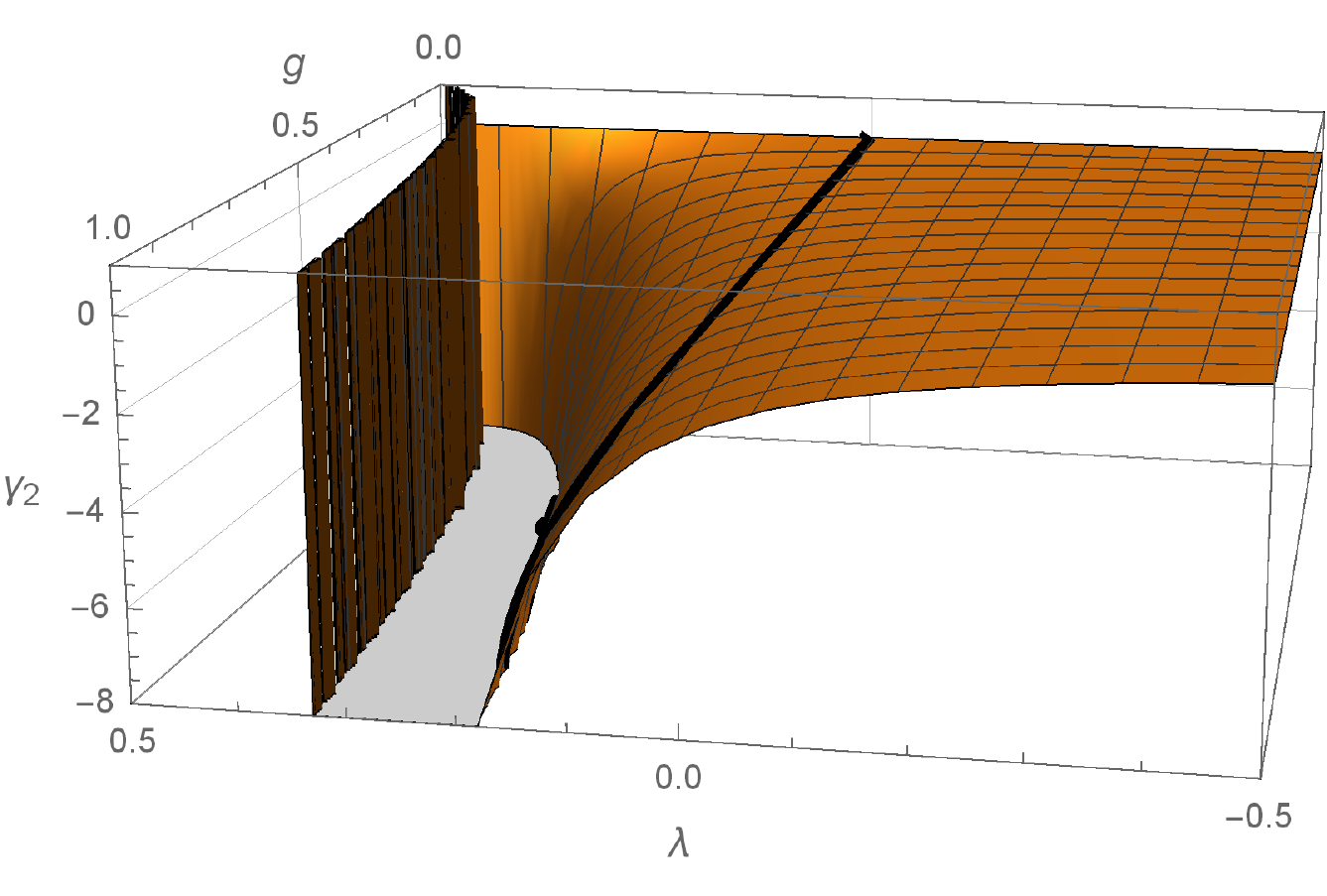} 
	\caption{\label{Fig.2} Illustration of the anomalous dimensions $\gamma_n$ for $n=0,1,2$ (top, middle, bottom row) and $d=3$ (left column) and $d=4$ (right column) as a function of $g$ and $\lambda$. The GFP and NGFP are marked with dots while the distinguished RG trajectory connecting the fixed points is shown as the black line.}
\end{figure}

\begin{figure}[t]
	\centering
	\includegraphics[width=0.48\textwidth]{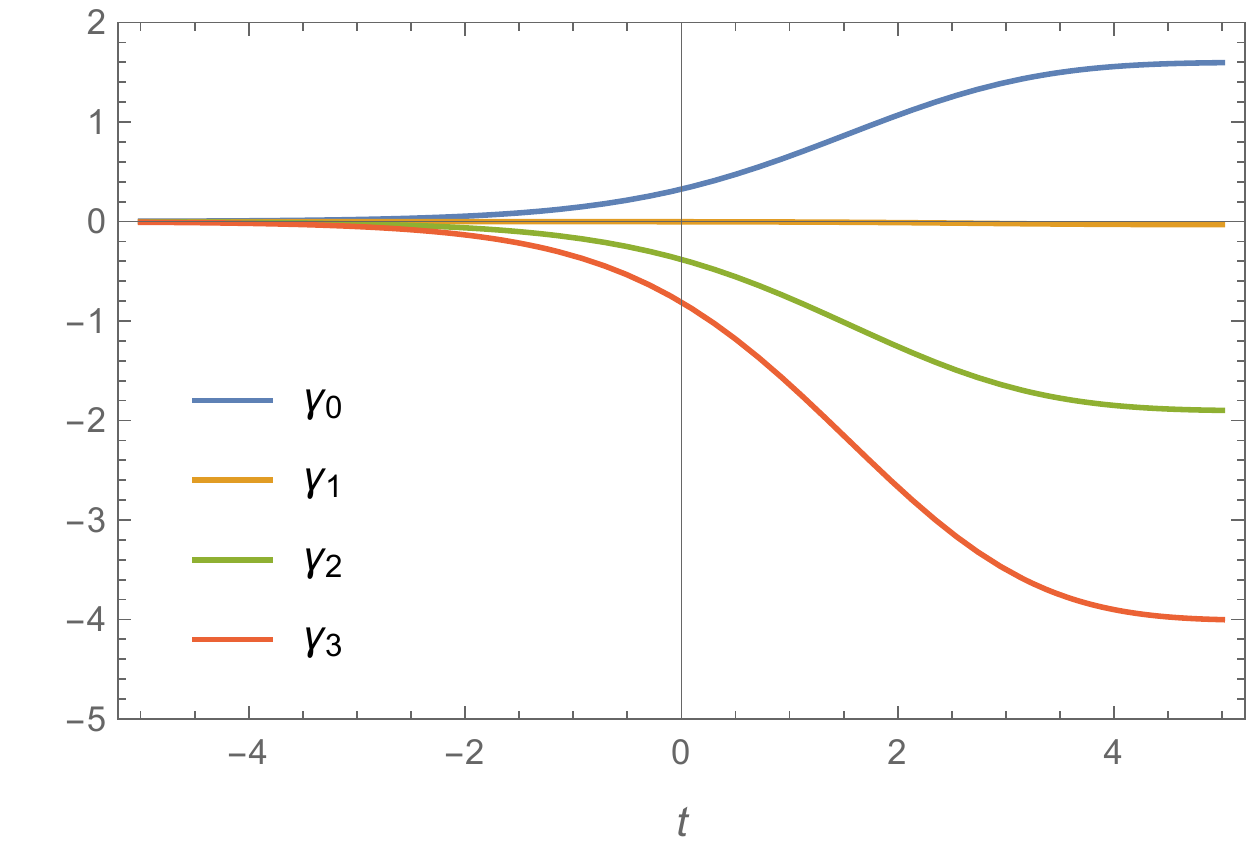}
	\includegraphics[width=0.48\textwidth]{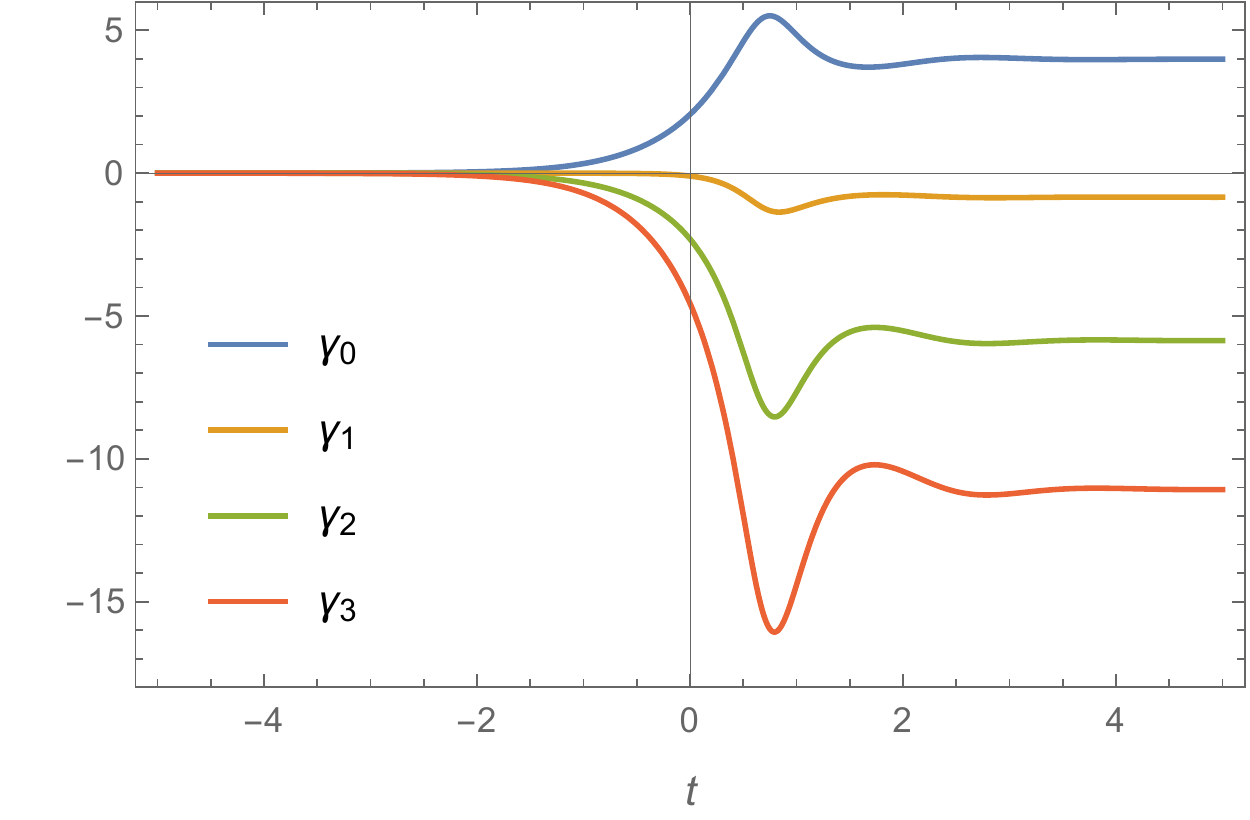}
	\caption{\label{Fig.4} Selected anomalous dimensions $\gamma_n$ evaluated along the renormalization group trajectory connecting the non-Gaussian to the Gaussian fixed point in $d=3$ (left) and $d=4$ (right). From top to bottom the lines correspond to $n=0,1,2,3$, respectively. The anomalous dimensions interpolate between their values at the NGFP given in \eqref{ScalingMaster} and zero in the limit $k\rightarrow 0$.}
\end{figure}

For $d=3$ and $d=4$ the value of the anomalous dimensions $\gamma_0, \gamma_1$ and $\gamma_2$ for
positive values of Newton's coupling $g$ are illustrated in Figure \ref{Fig.2}.
As a key feature, we observe that all $\gamma_n$ diverge at the locus where $\eta_N$ becomes infinite, c.f.\ Figure \ref{Fig.1}. For values $\lambda$ smaller than this singular line, we have
$\gamma_0 \ge 0$ while the higher-order values of $n$ yield $\gamma_1, \gamma_2, \cdots \leq 0$. 
Fig.\ \ref{Fig.2} also indicates the position of the fixed points (black dots) as well as the RG trajectory connecting them (black line). The ``flow'' of $\gamma_n$ along this trajectory is displayed in Fig.\ \ref{Fig.4}.
The figure illustrates the cross-over behavior of the $\gamma_n$: when $k$ is larger than the crossover scale $k_0$ ($t >0$) the geometry is characterized by the non-trivial scaling of the geometric operators dictated by the fixed point relations \eqref{ScalingMaster}. At $k=0$ all quantum effects vanish and one recovers a classical geometry. The RG flow then interpolates between these two regimes.

\begin{figure}[t]
	\centering
	\includegraphics[width=0.8\textwidth]{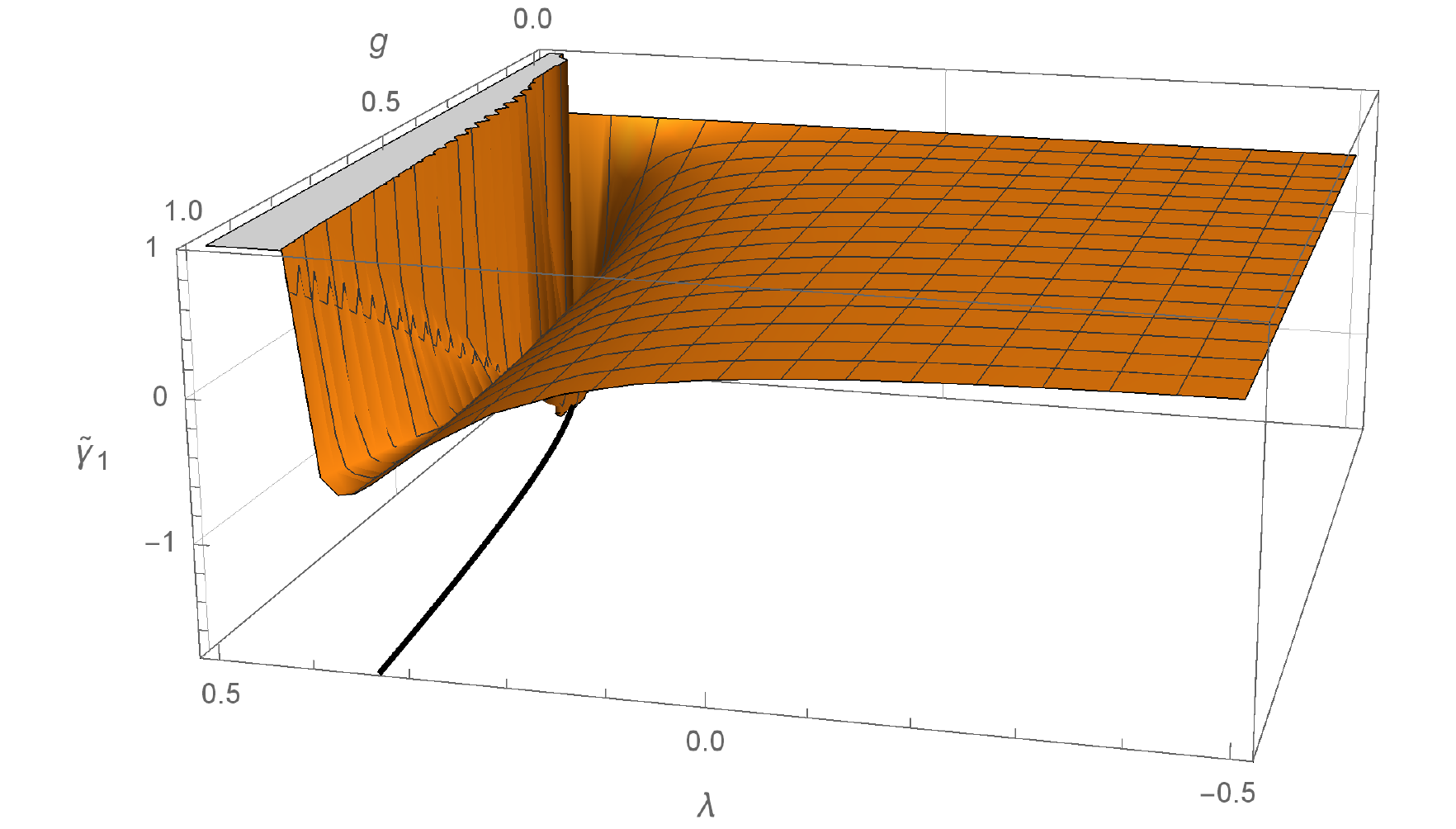}
	\caption{\label{Fig.5} Illustration of the scaling ratio \eqref{gtilde} evaluated for $d=4$ on the $g$-$\lambda$--plane. The black line in the bottom plane illustrates the position of the singular locus created by the divergence of $\eta_N$, cf.\ Fig.\ \ref{Fig.1}.}
\end{figure}

We close this subsection by illustrating the implications of our main result \eqref{mainresult} for the redefined scaling exponents \eqref{scalingexp}. For concreteness we chose the reference operator as the spacetime volume and focus on the specific combination
\be\label{gtilde}
\tilde{\gamma}_1 \equiv \frac{d_1 + \gamma_1}{d_0 + \gamma_0} = \frac{d-2+ \gamma_1}{d+\gamma_0} \, , 
\ee
in $d=4$ dimensions. We stress that \eqref{mainresult} allows for the easy construction of any scaling ratio in arbitrary dimension $d$ though. The values of $\tilde{\gamma}_1$ at the NGFP \eqref{NGFPpos} and the GFP are
\be
\tilde{\gamma}_1^{\rm NGFP} = 0.145 \, , \qquad \tilde{\gamma}_1^{\rm GFP} = \frac{1}{2} \, . 
\ee
The later is determined completely from the classical scaling dimensions of the operators $\cO_n$ and follows from \eqref{gtilde} by using that the anomalous dimensions vanish at the GFP. The value of $\tilde{\gamma}_1$ as a function of $g$ and $\lambda$ is illustrated in Fig.\ \ref{Fig.5}. 
The figure indicates that $\tilde{\gamma}_1$ differs appreciably from its classical value for $\lambda \gtrsim 0.2$ only. Remarkably, its value \emph{remains finite} at the locus where $\eta_N$ exhibits a divergence. The bottom of the valley runs close to (but does not agree with) this singular locus. For $\lambda=1/2$, $\tilde{\gamma}_1$ diverges. This divergence is caused by the graviton propagators derived from the Einstein-Hilbert action \eqref{EHprop} which are singular at this line.

%-----------------------------------------------------
\subsection{Comparison to the polynomial $f(R)$-truncation}
\label{sect.3.4}
%-----------------------------------------------------
At this stage, it is instructive to compare the results \eqref{ScalingMaster} to the ones obtained from solving the Wetterich equation \eqref{FRGE} projected onto gravitational actions of $f(R)$-type \cite{Codello:2007bd,Machado:2007ea,Benedetti:2012dx}. Performing a polynomial expansion of $f(R) \simeq \sum_{n=0}^N \bar{u}_n \, R^n$, the eigenvalues of the stability matrix \eqref{stabmat} have been determined for successively increasing values $N=6$ \cite{Codello:2007bd,Machado:2007ea}, $N=8$ \cite{Codello:2008vh}, $N=35$ \cite{Falls:2013bv,Falls:2014tra}, and lately also $N=71$ \cite{Falls:2018ylp}. Building on previous investigations \cite{Falls:2013bv,Falls:2014tra}, ref.\ \cite{Falls:2018ylp} reported that for large values of $n$ the real parts of the eigenvalues $\lambda_n$ follow an almost Gaussian behavior

\be\label{fRfit}
\lambda_n^{f(R)} \approx  a \, n - b  \, ,
\ee
where the best-fit values for the parameters $a$ and $b$ are 
\be\label{evest}
a = 2.042 \pm 0.002 \, , \qquad b = 2.91 \pm 0.05 \, . 
\ee

\begin{figure}[t]
	\centering
	\includegraphics[width=0.48\textwidth]{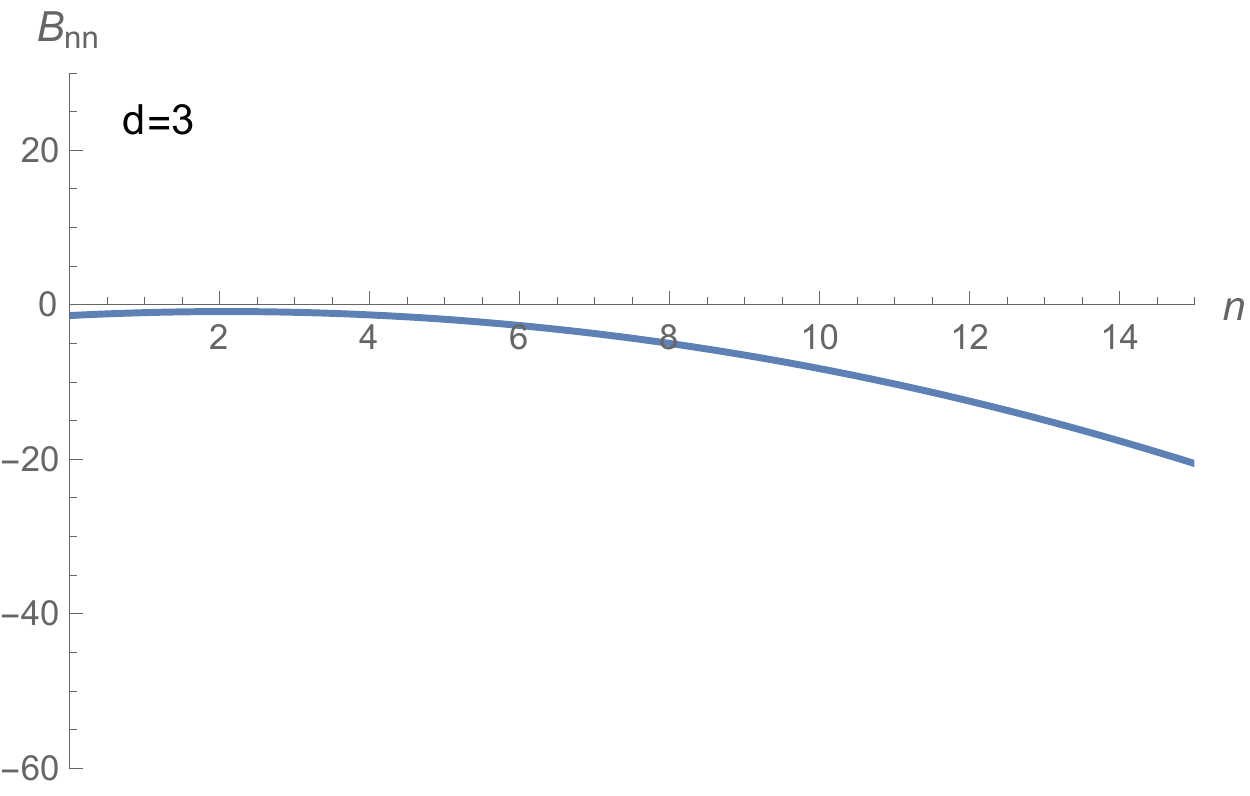}
	\includegraphics[width=0.48\textwidth]{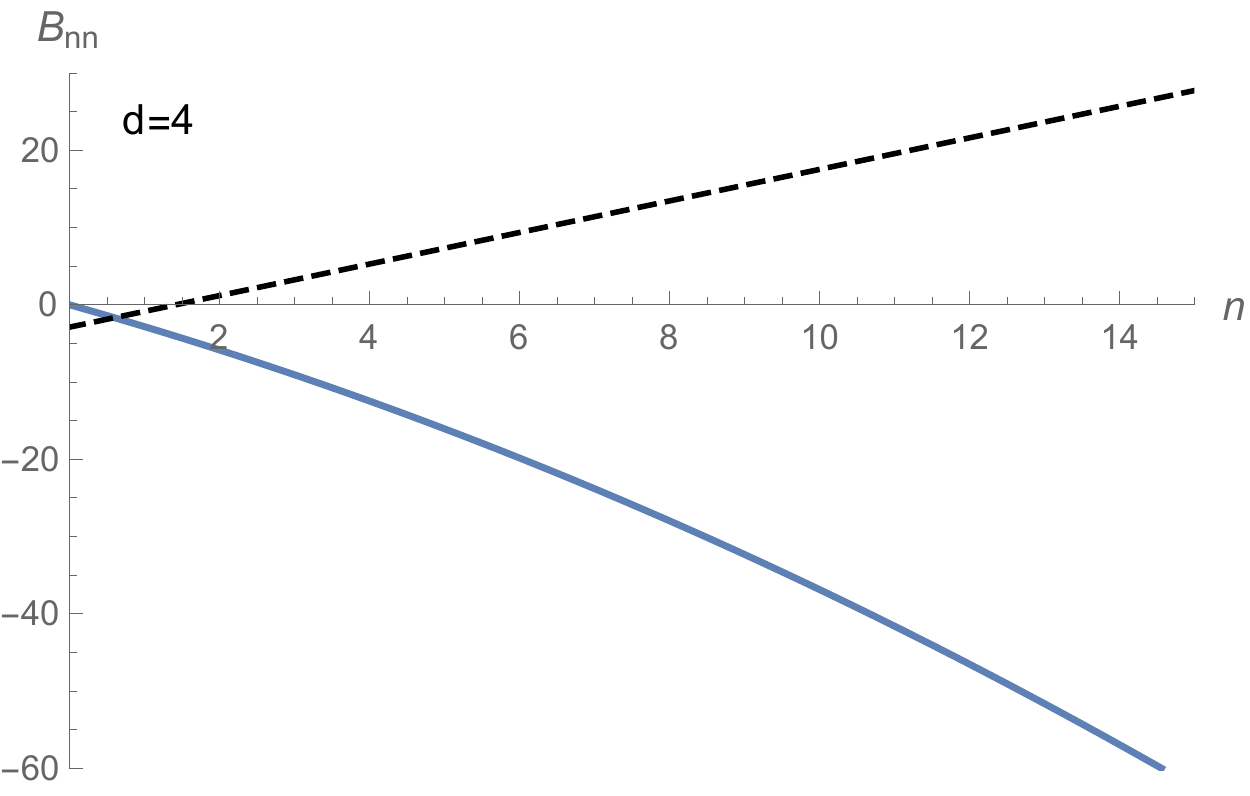}
	\caption{\label{Fig.6} Value of the diagonal matrix elements \eqref{Bnneq} as a function of $n$ for $d=3$ (left) and $d=4$ (right). For comparison the eigenvalue estimate \eqref{fRfit} of the stability matrix in $d=4$ has been added as a dashed line.}
\end{figure}

In order to connect the results \eqref{fRfit} and \eqref{ScalingMaster} we follow the strategy outlined in \cite{Pagani:2016dof}. Consider an ansatz for the effective average action,
\be
\Gamma_k = \bar{\Gamma}_k + \epsilon \, \ub \, \cO \, ,
\ee
where we have singled out one specific ($k$-independent) operator $\cO$ and introduced a dimensionless control parameter $\epsilon$. Substituting this ansatz into the Wetterich equation \eqref{FRGE} and considering the equation resulting at order $\epsilon$ gives
\be\label{dub}
\p_t \ub \, \cO = - \frac{1}{2} \left. {\rm Tr} \left[ \left(\bar{\Gamma}_k^{(2)} + \cR_k \right)^{-1} \, \ub \, \cO^{(2)} \, \left(\bar{\Gamma}_k^{(2)} + \cR_k \right)^{-1} \, \p_t \cR_k       \right] \right|_{\cO} \, . 
\ee
Assuming that $\cO$ has a canonical mass dimension $[\cO] = - d_u$, we can re-express $\ub$ in terms of the corresponding dimensionless coupling $u = \ub \, k^{-d_u}$
\be\label{dub2}
(\p_t u + d_u \, u) \, \cO = - \frac{1}{2} \left. {\rm Tr} \left[ \left(\bar{\Gamma}_k^{(2)} + \cR_k \right)^{-1} \, u \, \cO^{(2)} \, \left(\bar{\Gamma}_k^{(2)} + \cR_k \right)^{-1} \, \p_t \cR_k       \right] \right|_{\cO} \, . 
\ee
The trace appearing on the right-hand side is the one entering into the defining equation for the anomalous dimension of the operator $\cO$, eq.\ \eqref{CompositeMaster}. Thus, we arrive at the simple equation
\be
\p_t u = (\gamma_u - d_u) \, u \, , 
\ee
where $\gamma_u$ and $d_u$ are independent of $u$. Recalling the definitions $\p_t u = \beta_u$ and ${\bf B}_{uu} = \frac{\p \beta_u}{\p u}$, then leads to a simple relation between the diagonal matrix element of the stability matrix and the anomalous dimension of $\cO$:
\be
{\bf B}_{uu} = - d_u + \gamma_u \, . 
\ee
For vanishing quantum corrections $\gamma_u = 0$, this relation entails that the ${\bf B}_{uu}$ are given by minus the mass dimension of the coupling. For the operators \eqref{C} these are given by $d_n = d - 2n$. Using the anomalous dimensions \eqref{ScalingMaster}, one finds that the diagonal entries of the stability matrix associated with the operators $\cO_n$ are
\be\label{Bnneq}
\begin{split}
	d=3: \quad {\bf B}_{nn} = & -1.409 + 0.495 n - 0.118 n^2 \, , \\
	d=4: \quad {\bf B}_{nn} = & -0.013 - 2.733 n - 0.095 n^2 \, . 
\end{split}
\ee
Notably, the ${\bf B}_{nn}$ are negative for all values $n$. The corresponding polynomials are illustrated in Figure \ref{Fig.6}. 
The comparison with \eqref{fRfit} furthermore shows that there is no relation between the ${\bf B}_{nn}$ and the eigenvalues of the stability matrix. This is a strong indication that the eigenvalues of the later are actually fixed by the off-diagonal terms in ${\bf B}_i{}^j$ which can not be accessed from the single-operator approximation employed in \eqref{CompositeMaster}. While the difference between \eqref{Bnneq} and \eqref{fRfit} is not a contradiction, it still indicates that the composite operator formalism in this simple implementation is not able to determine the number of relevant directions at the NGFP. We will come back to this point at the end of the next section.

%-----------------------------------------------------
\section{Summary and discussion}
\label{sect.4}
%-----------------------------------------------------
The characterization of fluctuating geometries is one of the central open problems in quantum gravity research. Its importance can hardly be understated since ultimately this avenue may allow comparing predictions from currently distinct approaches to quantum gravity.\footnote{For earlier work along these lines comparing the spectral dimension and reconstructing the quantum effective action see \cite{Reuter:2011ah} and \cite{Ambjorn:2008wc,Ambjorn:2016fbd,Knorr:2018kog}, respectively.} In this work, we took an important next step in this direction by computing the anomalous scaling dimension of an infinite number of geometric operators. It is expected that Monte Carlo simulations (e.g., within the framework of Causal Dynamical Triangulations \cite{Ambjorn:2012jv,Loll:2019rdj}) these quantities is able to access these quantities more easily than, e.g., the critical exponents associated with the non-Gaussian fixed point underlying Asymptotic Safety. 

The main result of our work is the relation \eqref{mainresult}. This formula assigns an infinite set of anomalous scaling dimensions $\gamma_n$ to every point in the $g$-$\lambda$--plane. Notably this assignment is independent of any specific renormalization group trajectory or the existence of a fixed point. The values of $\gamma_n$ characterizing the non-Gaussian fixed points are found by evaluating this map at the location of the fixed point $\{\lambda_*,g_*\}$. Given an (approximate) solution of the Wetterich equation as, e.g., the one shown in Fig.\ \ref{Fig.1}, allows to trace the change in the  
anomalous dimensions when integrating out quantum fluctuations. For renormalization group trajectory connecting the non-Gaussian fixed point to the classical low-energy regime all anomalous dimensions vanish as $k \rightarrow 0$. This serves as a strong indicator that one recovers a classical geometry at low energy.

A rather remarkable feature of the map \eqref{mainresult} is that it \emph{does not change in the presence of matter fields}. This follows from the observation that the geometric operators \eqref{C} do not couple to matter sectors contained in $\Gamma_k$. Nevertheless, non-Gaussian fixed points associated to specific gravity-matter systems will come with their own characteristic set of anomalous dimensions: the matter content shifts the position of the non-Gaussian fixed point in the $g$-$\lambda$--plane, so that the map is evaluated at different values $(\lambda_*,g_*)$.

At this point the following critical remark is in order. In this work we explicitly demonstrated that the composite operator formalism based on the effective average action allows to derive \emph{ analytic results} for the anomalous scaling dimension of an infinite class of geometric operators. The comparison to similar results based on the stability matrix obtained by solving the Wetterich equation for $f(R)$-gravity showed that the anomalous dimensions are unrelated to the eigenvalues of the stability matrix. Hence our result can not be used to estimate the number of relevant parameters associated with the non-Gaussian fixed point. This is not a contradiction though and the reason can be understood rather easily. Inspecting the composite operator equation \eqref{CompositeMaster}, one finds that the operators \eqref{C} are not eigenoperators of the equation. In fact, expanding the right-hand side obtained for the operator $\cO_n = \int d^dx \sqrt{g} R^n$, $n \ge 2$ on a spherical symmetric background generates an infinite tower of operators $\cO_{n-2}, \cO_{n-1}, \cdots$. Thus the eigenoperators are linear combinations of the $\cO_n$ involving an infinite number of terms. It is then clear that the composite operator equation in its one-dimensional approximation determines the diagonal matrix elements of the stability matrix only. In order to determine its eigenvalues knowledge about the off-diagonal terms is required. The example provided by the Einstein-Hilbert truncation given in eq.\ \eqref{stabmatex} indicates that these terms provide a significant contribution and can not be ignored when determining the relevant parameters of the non-Gaussian fixed point. Thus applying the composite operator formalism to determine the relevant parameters requires a priori knowledge on the basis of geometric operators which diagonalize the operator trace on its right-hand side. The present result can be understood as a first step in this direction, illustrating that such a computation is actually feasible (at least on a spherically symmetric background). We leave this investigation for future work though. 

\appendix
%-----------------------------------------------------
\section{Computing traces with non-minimal operator insertions}
\label{App.A}
%-----------------------------------------------------
The evaluation of the operator traces \eqref{mastertraces} is most conveniently done by applying the early-time expansion of the heat-kernel in combination with Laplace-transform techniques \cite{Codello:2008vh,Reuter:2019byg}. The underlying technical details are collected in this appendix.
%-----------------------------------------------------
\subsection{Early-time heat-kernel and Mellin transforms}
%-----------------------------------------------------
Inspecting the arguments of the traces appearing in eq.\ \eqref{mastertraces} one finds that they already contain at least $n-2$ powers of the background curvature $\Rb$. When performing the projection onto the geometric operators \eqref{C}, it suffices to have the early-time expansion of the heat-kernel for scalars (S) to second order in the curvature while for traceless tensors (T) and transverse vectors (VT) the result to first order is sufficient. Evaluating the general expression \cite{Vassilevich:2003xt} for the spherically symmetric background \eqref{background} yields
\ba\label{heatkernel}
{\rm Tr}_i \left[e^{-s\Delta}\right] \! \! \! \! & \simeq & \! \! \! \! \frac{1}{(4\pi s)^{d/2}} \int d^dx \sqrt{\gb} \left[ a_0^i + a_1^i \, s \, \Rb + a_2^i \, s^2 \, \Rb^2 + \ldots\right] \, ,
\ea
where ``$i$'' labels the type of field the Laplacian acts on and the dots represent higher-order curvature terms which do not enter in the present computation. The coefficients $a_n^i$ required in the computation are listed in Table \ref{Tab.1}.
\begin{table}[t]
	\begin{center}
		\begin{tabular}{|c||c|c|c|}
			& S & T & TV \\ \hline \hline
			$a_0^i$ & $1$ & $\frac{(d+2)(d-1)}{2} $ & $d-1$ \\
			$a_1^i$ & $\frac{1}{6}$ & \quad $\frac{(d+2)(d-1)}{12}$ \quad & \quad $\frac{(d+2)(d-3)}{6d}$ \quad \\
			$a_2^i$ & \quad $\frac{5d^2-7d+6}{360 d (d-1)}$ \quad & $-$ & $-$\\ \hline \hline 
		\end{tabular}
		\caption{\label{Tab.1} Heat-kernel coefficients for scalars (S), traceless symmetric matrices (T), and transverse vectors (TV) on a background $d$-sphere \cite{Lauscher:2001ya}. The dash $-$ indicates that the corresponding coefficient is not entering into the present computation.}
	\end{center}
\end{table}

The heat-kernel formula \eqref{heatkernel} is readily generalized to functions of the Laplacian \cite{Codello:2008vh,Reuter:2019byg}
\be\label{heat2}
\begin{split}
	{\rm Tr}_i\left[W(\Delta) \right] = & \, \frac{1}{(4\pi)^{d/2}} \int d^dx \sqrt{\gb} \, \big[ a_0^i \, Q_{d/2}[W] + a_1^i \, Q_{d/2-1}[W] \, \Rb \\
	& \qquad \qquad \qquad  + a_2^i \, Q_{d/2-2}[W] \, \Rb^2 + \ldots \big] \, . 
\end{split}
\ee  
For $n > 0$, the Q-functionals are given by the Mellin-transform of $W(z)$,
\be
Q_n[W] = \frac{1}{\Gamma(n)} \int_0^\infty dz z^{n-1} W(z) \, , \qquad n > 0 \, , 
\ee
while for $n=0$ one has
\be
Q_0[W] = W(0) \, . 
\ee 
%-----------------------------------------------------
\subsection{Computing the master traces}
%-----------------------------------------------------
We now compute the operator traces determining the anomalous dimension of the geometric operators \eqref{C}. We start with the scalar sector and subsequently determine the minimal and non-minimal contributions from the traceless fluctuations. All dimensionless threshold functions are evaluated at $-2\lambda$ and we suppress the argument  in order to lighten the notation, i.e.\ $q^p_n \equiv q^p_n(-2\lambda)$.

\noindent
\emph{Scalar traces.} The scalar contribution is contained in the three master traces
\be
\begin{split}
	T_0^S = & \, {\rm Tr_S}\left[W^S(\Delta)\right]|_{\Rb^0} \, , \\
	T_1^S = & \, {\rm Tr_S}\left[\Delta \, W^S(\Delta)\right]|_{\Rb^1} \, , \\
	T_2^S = & \, {\rm Tr_S}\left[\Delta^2 \, W^S(\Delta)\right]|_{\Rb^2} \, . \\
\end{split}
\ee
Here $|_{\Rb^n}$ indicates that only terms containing $n$-powers of the background scalar curvature are relevant for the computation. Abbreviating
\be
N_S \equiv - 32 \pi g_k \, \frac{2d}{d-2} \, , 
\ee
we get
\be
\begin{split}
	T_0^S = & \frac{2 N_S}{(4\pi)^{d/2}} \, q^2_{d/2} \, \int_x	\, , \\
	T_1^S = & \frac{N_S}{(4\pi)^{d/2}} \, \left(a_1^S \, (d-2) \, q^2_{d/2} - 2 d \, C_S \, q^3_{d/2+1}\right) \int_x \, \Rb \, , \\
	T_2^S = & \frac{N_S}{(4\pi)^{d/2}}  \Big[ 
	\frac{3(d+2)d}{2} \, C_S^2 \, q^4_{d/2+2} 
	- a_1^S d(d-2) C_S \, q^3_{d/2+1} \\ & \qquad \qquad
	+ \frac{1}{2} \, a_2^S \, (d-2)(d-4) \, q^2_{d/2} 
	\Big] \, \int_x \Rb^2\, . 
\end{split}
\ee

\noindent
\emph{Tensor traces I - minimal terms.}
Inspecting the terms proportional to $\mathbbm 1$ in eq.\ \eqref{Ohtht}, the relevant master traces in the tensor sector are
\be\label{A.10}
\begin{split}
	T_0^T = & \, {\rm Tr_T}\left[W^T(\Delta)\right]|_{\Rb^0} \, , \\
	T_1^T = & \, {\rm Tr_T}\left[\Delta \, W^T(\Delta)\right]|_{\Rb^1} \, . 
\end{split}
\ee
Following the computation in the scalar sector, these evaluate to
\be
\begin{split}
	T_0^T = & \frac{64 \pi g}{(4\pi)^{d/2}} \, a_0^T \, q^2_{d/2} \, \int_x	\, , \\
	T_1^T = & \frac{32 \pi g}{(4\pi)^{d/2}} \, \left[a_1^T \, (d-2) \, q^2_{d/2} - 2 d \, a_0^T \, C_T \, q^3_{d/2+1}\right] \int_x \, \Rb \, .
\end{split}
\ee

\noindent
\emph{Tensor traces II - non-minimal terms.}
In addition to the traces over Laplacian-valued functions \eqref{A.10}, the tensor sector gives rise to two additional traces including operator insertions of non-Laplace type
\be\label{A.12}
\begin{split}
	T_2^T = & \, {\rm Tr_T}\left[ \delta^{(\mu}_{(\alpha} \Db^{\nu)} \Db_{\beta)} \, W_T(\Delta)\right]|_{\Rb^1} \, , \\
	T_3^T = & \, {\rm Tr_T}\left[\Db^{(\mu} \Db^{\nu)} \Db_{(\alpha} \Db_{\beta)} \, W_T(\Delta)\right]|_{\Rb^2} \, . \\
\end{split}
\ee
Owed to the zero-modes associated with the operator insertion the computation of the traces is slightly more involved. We start by further decomposing the fluctuation field $\hh_{\mu\nu}$ into its transverse-traceless part $h^{\rm TT}_{\mu\nu}$, a transverse vector $\xi^\mu$ and a scalar $\sigma$ \cite{York:1973ia,Benedetti:2010nr}
\be\label{TTdec}
\hh_{\mu\nu} = h^{\rm TT}_{\mu\nu} + \Db_\mu \xi_\nu + \Db_\nu \xi_\mu + \left( \Db_\mu \Db_\nu - \frac{1}{d} \gb_{\mu\nu} \Db^2 \right) \sigma \, , 
\ee 
where the component fields are subject to the differential constraints
\be
\gb^{\mu\nu} h^{\rm TT}_{\mu\nu} = 0 \, , \qquad \Db^\mu h^{\rm TT}_{\mu\nu} = 0 \, , \qquad \Db_\mu \xi^\mu = 0 \, . 
\ee
The Jacobian associated with the field redefinition \eqref{TTdec} can be set to one by redefining the fields according to 
\be
\sqrt{2} \left[\Delta - \frac{1}{d} \Rb \right]^{1/2} \xi_\mu \mapsto  \, \xi_\mu \, , \qquad \left[ \frac{d-1}{d} \Delta^2 - \frac{1}{d} \Rb \Delta \right]^{1/2} \sigma \mapsto  \sigma \, . 
\ee
Performing the transverse-traceless decomposition at the level of the traces, the operators $\left[\left. \cO^{(2),{\rm nm}}_n \right|_{\hh\hh}\right]^{\mu\nu}{}_{\alpha\beta}$ annihilate the terms including the transverse-traceless part of the fluctuation field. Furthermore,  the non-minimal traces \eqref{A.12} reduce to traces involving Laplace-type operators only. Explicitly, the transverse-traceless decomposition of $T_2^T$ yields
\be\label{T2Tprev}
\begin{split}
	T_2^T = & \, - \frac{1}{2} {\rm Tr}_{\rm TV}\left[ \left(\Delta - \tfrac{1}{d} \Rb \right) W_T(\Delta) \right] + \frac{d+1}{2d(d-1)}\, \Rb  \, {\rm Tr}_{\rm TV}\left[ \Delta  W^{\prime}_T(\Delta) \right] \\
	& \, - {\rm Tr}_{0}\left[\left( \tfrac{d-1}{d} \Delta - \tfrac{1}{d} \Rb \right)  W_T(\Delta) \right] + \frac{2}{d} \, \Rb \, {\rm Tr}_{0}\left[ \Delta \, W_T^{\prime}(\Delta) \right] \, ,
\end{split}
\ee
where the prime denotes a derivative with respect to the argument and terms of order $\Rb^2$ are dropped.  Evaluating the terms at order $\Rb$ using the early-time expansion of the heat-kernel the relevant contribution stemming from $T^T_2$ is
\be
\begin{split}
	T^T_2 = & \, - \frac{32 \pi g}{(4\pi)^{d/2}} \Big[\frac{(d+2)^2(d+1)}{12d} \, q^2_{d/2} - (d+2)(d-1) \, C_T \, q^3_{d/2+1}\Big] \int_x \, \Rb .
\end{split}
\ee
Note that the derivatives of $W_T$ can be converted to standard $q$-functionals by performing a  partial integration within the Q-functional.

The transverse-traceless decomposition of $T^T_3$ results in
\be\label{T3Tprev}
\begin{split}
	T^T_3 = & \, {\rm Tr}_0 \left[ \left( \tfrac{d-1}{d}\Delta^2 - \tfrac{1}{d} \Rb \Delta \right) \, W_{T}(\Delta) \right] 
	- \frac{2}{d-1} \Rb \, {\rm Tr}_0 
	\left[ \left( \tfrac{d-1}{d}\Delta^2 - \tfrac{1}{d} \Rb \Delta \right) \, W^{\prime}_T(\Delta)  \right] \\ & \, 
	+ \frac{2}{d(d-1)} \, \Rb^2 \, {\rm Tr}_0 
	\left[ \Delta^2 \, W_T^{\prime\prime}(\Delta)  \right] \, , 
\end{split}
\ee
where we dropped terms of order $\Rb^3$. Labeling the contribution in the first, second, and third line by $T_{3a}^T$, $T_{3b}^T$, and $T_{3c}^T$, respectively, we note the following intermediate results
\be
\begin{split}
	T_{3a}^T = & \frac{32 \pi g}{(4\pi)^{d/2}}  \Big[ 
	\frac{3(d-1)(d+2)}{2} \, C_T^2 \, q^4_{d/2+2} 
	- \frac{ (d-5)(d+2)}{6} C_T \, q^3_{d/2+1} \\ & \qquad \qquad
	+ \frac{(d+2)(d-2)(5d^2-37d-12)}{720 d^2} \, q^2_{d/2} 
	\Big] \, \int_x \Rb^2\, . \\
	T_{3b}^T = &  \frac{32 \pi g}{(4\pi)^{d/2}}  \left[ \frac{(d-5)(d+2)}{6(d-1)}
	\, q^2_{d/2} - 2 (d+2) C_T \, q^3_{d/2+1}
	\right]
	\, \int_x \Rb^2 \, . \\
	T_{3c}^T = & \frac{32 \pi g}{(4\pi)^{d/2}}  \, \frac{d+2}{d-1}  \, q^2_{d/2} \, \int_x \, \Rb^2 . 
\end{split}
\ee
where we display the terms at the relevant order of $\Rb$. Adding these terms gives the final result for $T^T_3$:
\be
\begin{split}
	T_3^T = & \, \frac{32 \pi g}{(4\pi)^{d/2}}  \Big[ 
	\frac{3}{2} (d+2)(d-1)\, C_T^2 \, q^4_{d/2+2}
	- \frac{1}{6} (d+7) (d+2) 
	C_T \, q^3_{d/2+1} \\ & \qquad
	+\frac{5d^5 + 78 d^4  +365 d^3 + 420 d^2 -100 d - 48}{720 d^2 (d-1)}
	\, q^2_{d/2}
	\Big] \, \int_x \Rb^2\, .
\end{split}
\ee
This concludes the derivation of the intermediate results contributing to \eqref{mainresult}.
%-----------------------------------------------------
\section{Commutator relations on the $d$-sphere}
\label{App.B}
%-----------------------------------------------------
Evaluating the traces including a non-minimal operator insertion requires 
commuting covariant derivatives with functions of the Laplacian. The structure of \eqref{mastertraces} indicates that it suffices to evaluate the commutator relations up to order $R^2$. The relevant formulas are given in App.\ A of Ref.\ \cite{Benedetti:2010nr} and we include them here for completeness. 

Assuming that the function $W(\Delta)$ can be represented by an (inverse) Laplace-transform, one readily proofs
\be\label{comfct}
\begin{split}
	\Big[ & \Db_\mu , W(\Delta)\Big]  \, \phi_{\alpha_1 \cdots \alpha_n} = \\ & \,
	\left\{W^\prime(\Delta) [\Db_\mu, \Delta] + \tfrac{1}{2} W^{\prime\prime}(\Delta) \left[ \left[ \Db_\mu, \Delta \right], \Delta \right] \right\} \phi_{\alpha_1 \cdots \alpha_n} + \cO(\Db \Rb, \Rb^3) \, , 
\end{split}
\ee
where the right-hand side is exact up to terms containing derivatives of the curvature tensor and terms involving three curvature tensors. 

The derivation of \eqref{T2Tprev} and \eqref{T3Tprev} based on \eqref{comfct} then uses the following commutator relations, valid on a background $d$-sphere
\ba\label{com1a}
\left[ \Db_\mu, \Delta \right] \hh^{\mu\nu} = & \!\!\!\!\! - \frac{d+1}{d(d-1)} \Rb \Db_\mu \hh^{\mu\nu} \, , \\ \left[ \Db_\mu, \Delta \right] \phi^{\mu} = & \!\!\!\!\! \!\!\!\!\! \!\!\!\!\! - \frac{1}{d} \Rb \Db_\mu \phi^\mu \, .
\ea
Here $\hh^{\mu\nu}$ is symbolic for any traceless, symmetric tensor. Using \eqref{com1a}, it is also straightforward to evaluate the double-commutator
\be
\left[\left[ \Db_\mu, \Delta \right], \Delta \right] \hh^{\mu\nu} = \frac{(d+1)^2}{d^2(d-1)^2} \Rb^2 \Db_\mu \hh^{\mu\nu} \, . 
\ee

%-----------------------------------------------------
\acknowledgments

 We thank C.\ Pagani, J.\ Pawlowski and M.\ Reuter for useful discussions.  Moreover, we are greatful to A.\ Barvinsky, J.\ Daas and N.\ Klitgaard for a careful reading of the manuscript and many valuable comments. F.S.\ is acknowledges financial support by the Netherlands Organisation  for  Scientific  Research  (NWO)  within  the  Foundation  for  Fundamental Research on Matter (FOM) grant 13VP12. The work of A.K.\ has been supported by the Foundation for Theoretical Physics Development Basis, the RFBR grant No. 20-02-00297, as well as an ERASMUS+ fellowship.
%-----------------------------------------------------


\begin{thebibliography}{99}

\bibitem{Weinberg:1976xy}
S.~Weinberg, {Critical Phenomena for Field Theorists}, in: {Erice
	Subnucl.Phys.1976:1}, 1976.

\bibitem{Weinberg:1980gg}
S.~Weinberg, {Ultraviolet divergences in quantum theories of gravitation},
General Relativity: An Einstein centenary survey, Eds. Hawking, S.W., Israel,
W; Cambridge University Press (1979) 790.

\bibitem{Percacci:2017fkn}
R.~Percacci, {An Introduction to Covariant Quantum Gravity and Asymptotic
	Safety}, Vol.~3 of 100 Years of General Relativity, World Scientific, 2017.
\newblock \href {http://dx.doi.org/10.1142/10369} {\path{doi:10.1142/10369}}.

\bibitem{Reuter:2019byg}
M.~Reuter, F.~Saueressig, {Quantum Gravity and the Functional Renormalization
	Group}, Cambridge University Press, 2019.

\bibitem{Reuter:1996cp}
M.~Reuter, {Nonperturbative evolution equation for quantum gravity}, Phys.Rev.
D57 (1998) 971--985.
\newblock \href {http://arxiv.org/abs/hep-th/9605030}
{\path{arXiv:hep-th/9605030}}, \href
{http://dx.doi.org/10.1103/PhysRevD.57.971}
{\path{doi:10.1103/PhysRevD.57.971}}.

\bibitem{Wetterich:1992yh}
C.~Wetterich, {Exact evolution equation for the effective potential},
Phys.\ Lett.\ B301 (1993) 90.
\newblock \href {http://dx.doi.org/10.1016/0370-2693(93)90726-X}
{\path{doi:10.1016/0370-2693(93)90726-X}}.

\bibitem{Morris:1993qb}
T.~R. Morris, {The Exact renormalization group and approximate solutions}, Int.
J. Mod. Phys. A9 (1994) 2411.
\newblock \href {http://arxiv.org/abs/hep-ph/9308265}
{\path{arXiv:hep-ph/9308265}}, \href
{http://dx.doi.org/10.1142/S0217751X94000972}
{\path{doi:10.1142/S0217751X94000972}}.

\bibitem{Reuter:2001ag}
M.~Reuter, F.~Saueressig, {Renormalization group flow of quantum gravity in the
	Einstein-Hilbert truncation}, Phys. Rev. D65 (2002) 065016.
\newblock \href {http://arxiv.org/abs/hep-th/0110054}
{\path{arXiv:hep-th/0110054}}, \href
{http://dx.doi.org/10.1103/PhysRevD.65.065016}
{\path{doi:10.1103/PhysRevD.65.065016}}.

\bibitem{Souma:1999at}
W.~Souma, {Nontrivial ultraviolet fixed point in quantum gravity}, Prog. Theor.
Phys. 102 (1999) 181.
\newblock \href {http://arxiv.org/abs/hep-th/9907027}
{\path{arXiv:hep-th/9907027}}, \href {http://dx.doi.org/10.1143/PTP.102.181}
{\path{doi:10.1143/PTP.102.181}}.

\bibitem{Falkenberg:1996bq}
S.~Falkenberg, S.~D. Odintsov, {Gauge dependence of the effective average
	action in Einstein gravity}, Int. J. Mod. Phys. A13 (1998) 607.
\newblock \href {http://arxiv.org/abs/hep-th/9612019}
{\path{arXiv:hep-th/9612019}}, \href
{http://dx.doi.org/10.1142/S0217751X98000263}
{\path{doi:10.1142/S0217751X98000263}}.

\bibitem{Lauscher:2001ya}
O.~Lauscher, M.~Reuter, {Ultraviolet fixed point and generalized flow equation
	of quantum gravity}, Phys.Rev. D65 (2002) 025013.
\newblock \href {http://arxiv.org/abs/hep-th/0108040}
{\path{arXiv:hep-th/0108040}}, \href
{http://dx.doi.org/10.1103/PhysRevD.65.025013}
{\path{doi:10.1103/PhysRevD.65.025013}}.

\bibitem{Litim:2003vp}
D.~F. Litim, {Fixed points of quantum gravity}, Phys. Rev. Lett. 92 (2004)
201301.
\newblock \href {http://arxiv.org/abs/hep-th/0312114}
{\path{arXiv:hep-th/0312114}}, \href
{http://dx.doi.org/10.1103/PhysRevLett.92.201301}
{\path{doi:10.1103/PhysRevLett.92.201301}}.

\bibitem{Bonanno:2004sy}
A.~Bonanno, M.~Reuter, {Proper time flow equation for gravity}, JHEP 02 (2005)
035.
\newblock \href {http://arxiv.org/abs/hep-th/0410191}
{\path{arXiv:hep-th/0410191}}, \href
{http://dx.doi.org/10.1088/1126-6708/2005/02/035}
{\path{doi:10.1088/1126-6708/2005/02/035}}.

\bibitem{Eichhorn:2009ah}
A.~Eichhorn, H.~Gies, M.~M. Scherer, {Asymptotically free scalar
	curvature-ghost coupling in Quantum Einstein Gravity}, Phys. Rev. D80 (2009)
104003.
\newblock \href {http://arxiv.org/abs/0907.1828} {\path{arXiv:0907.1828}},
\href {http://dx.doi.org/10.1103/PhysRevD.80.104003}
{\path{doi:10.1103/PhysRevD.80.104003}}.

\bibitem{Manrique:2009uh}
E.~Manrique, M.~Reuter, {Bimetric Truncations for Quantum Einstein Gravity and
	Asymptotic Safety}, Annals Phys.\ 325 (2010) 785.
\newblock \href {http://arxiv.org/abs/0907.2617} {\path{arXiv:0907.2617}},
\href {http://dx.doi.org/10.1016/j.aop.2009.11.009}
{\path{doi:10.1016/j.aop.2009.11.009}}.

\bibitem{Eichhorn:2010tb}
A.~Eichhorn, H.~Gies, {Ghost anomalous dimension in asymptotically safe quantum
	gravity}, Phys. Rev. D81 (2010) 104010.
\newblock \href {http://arxiv.org/abs/1001.5033} {\path{arXiv:1001.5033}},
\href {http://dx.doi.org/10.1103/PhysRevD.81.104010}
{\path{doi:10.1103/PhysRevD.81.104010}}.

\bibitem{Groh:2010ta}
K.~Groh, F.~Saueressig, {Ghost wave-function renormalization in Asymptotically
	Safe Quantum Gravity}, J. Phys. A43 (2010) 365403.
\newblock \href {http://arxiv.org/abs/1001.5032} {\path{arXiv:1001.5032}},
\href {http://dx.doi.org/10.1088/1751-8113/43/36/365403}
{\path{doi:10.1088/1751-8113/43/36/365403}}.

\bibitem{Manrique:2010am}
E.~Manrique, M.~Reuter, F.~Saueressig, {Bimetric Renormalization Group Flows in
	Quantum Einstein Gravity}, Annals Phys. 326 (2011) 463.
\newblock \href {http://arxiv.org/abs/1006.0099} {\path{arXiv:1006.0099}},
\href {http://dx.doi.org/10.1016/j.aop.2010.11.006}
{\path{doi:10.1016/j.aop.2010.11.006}}.

\bibitem{Christiansen:2012rx}
N.~Christiansen, D.~F. Litim, J.~M. Pawlowski, A.~Rodigast, {Fixed points and
	infrared completion of quantum gravity}, Phys.Lett. B728 (2014) 114.
\newblock \href {http://arxiv.org/abs/1209.4038} {\path{arXiv:1209.4038}},
\href {http://dx.doi.org/10.1016/j.physletb.2013.11.025}
{\path{doi:10.1016/j.physletb.2013.11.025}}.

\bibitem{Codello:2013fpa}
A.~Codello, G.~D'Odorico, C.~Pagani, {Consistent closure of renormalization
	group flow equations in quantum gravity}, Phys.Rev. D89~(8) (2014) 081701.
\newblock \href {http://arxiv.org/abs/1304.4777} {\path{arXiv:1304.4777}},
\href {http://dx.doi.org/10.1103/PhysRevD.89.081701}
{\path{doi:10.1103/PhysRevD.89.081701}}.

\bibitem{Benedetti:2013jk}
D.~Benedetti, {On the number of relevant operators in asymptotically safe
	gravity}, Europhys. Lett. 102 (2013) 20007.
\newblock \href {http://arxiv.org/abs/1301.4422} {\path{arXiv:1301.4422}},
\href {http://dx.doi.org/10.1209/0295-5075/102/20007}
{\path{doi:10.1209/0295-5075/102/20007}}.

\bibitem{Becker:2014qya}
D.~Becker, M.~Reuter, {En route to Background Independence: Broken
	split-symmetry, and how to restore it with bi-metric average actions}, Annals
Phys. 350 (2014) 225.
\newblock \href {http://arxiv.org/abs/1404.4537} {\path{arXiv:1404.4537}},
\href {http://dx.doi.org/10.1016/j.aop.2014.07.023}
{\path{doi:10.1016/j.aop.2014.07.023}}.

\bibitem{Falls:2015qga}
K.~Falls, {On the renormalisation of Newton's constant}, Phys. Rev. D92 (2015)
124057.
\newblock \href {http://arxiv.org/abs/1501.05331} {\path{arXiv:1501.05331}},
\href {http://dx.doi.org/10.1103/PhysRevD.92.124057}
{\path{doi:10.1103/PhysRevD.92.124057}}.

\bibitem{Gies:2015tca}
H.~Gies, B.~Knorr, S.~Lippoldt, {Generalized Parametrization Dependence in
	Quantum Gravity}, Phys. Rev. D92 (2015) 084020.
\newblock \href {http://arxiv.org/abs/1507.08859} {\path{arXiv:1507.08859}},
\href {http://dx.doi.org/10.1103/PhysRevD.92.084020}
{\path{doi:10.1103/PhysRevD.92.084020}}.

\bibitem{Pagani:2016dof}
C.~Pagani, M.~Reuter, {Composite Operators in Asymptotic Safety}, Phys. Rev.
D95 (2017) 066002.
\newblock \href {http://arxiv.org/abs/1611.06522} {\path{arXiv:1611.06522}},
\href {http://dx.doi.org/10.1103/PhysRevD.95.066002}
{\path{doi:10.1103/PhysRevD.95.066002}}.

\bibitem{Falls:2017cze}
K.~Falls, {Physical renormalization schemes and asymptotic safety in quantum
	gravity}, Phys. Rev. D96 (2017) 126016.
\newblock \href {http://arxiv.org/abs/1702.03577} {\path{arXiv:1702.03577}},
\href {http://dx.doi.org/10.1103/PhysRevD.96.126016}
{\path{doi:10.1103/PhysRevD.96.126016}}.

\bibitem{Knorr:2017fus}
B.~Knorr, S.~Lippoldt, {Correlation functions on a curved background}, Phys.
Rev. D96 (2017) 065020.
\newblock \href {http://arxiv.org/abs/1707.01397} {\path{arXiv:1707.01397}},
\href {http://dx.doi.org/10.1103/PhysRevD.96.065020}
{\path{doi:10.1103/PhysRevD.96.065020}}.

\bibitem{Lauscher:2002sq}
O.~Lauscher, M.~Reuter, {Flow equation of quantum Einstein gravity in a higher
	derivative truncation}, Phys. Rev. D66 (2002) 025026.
\newblock \href {http://arxiv.org/abs/hep-th/0205062}
{\path{arXiv:hep-th/0205062}}, \href
{http://dx.doi.org/10.1103/PhysRevD.66.025026}
{\path{doi:10.1103/PhysRevD.66.025026}}.

\bibitem{Reuter:2002kd}
M.~Reuter, F.~Saueressig, {A Class of nonlocal truncations in quantum Einstein
	gravity and its renormalization group behavior}, Phys. Rev. D66 (2002)
125001.
\newblock \href {http://arxiv.org/abs/hep-th/0206145}
{\path{arXiv:hep-th/0206145}}, \href
{http://dx.doi.org/10.1103/PhysRevD.66.125001}
{\path{doi:10.1103/PhysRevD.66.125001}}.

\bibitem{Codello:2006in}
A.~Codello, R.~Percacci, {Fixed points of higher derivative gravity}, Phys.
Rev. Lett. 97 (2006) 221301.
\newblock \href {http://arxiv.org/abs/hep-th/0607128}
{\path{arXiv:hep-th/0607128}}, \href
{http://dx.doi.org/10.1103/PhysRevLett.97.221301}
{\path{doi:10.1103/PhysRevLett.97.221301}}.

\bibitem{Codello:2007bd}
A.~Codello, R.~Percacci, C.~Rahmede, {Ultraviolet properties of f(R)-gravity},
Int.J.Mod.Phys. A23 (2008) 143.
\newblock \href {http://arxiv.org/abs/0705.1769} {\path{arXiv:0705.1769}},
\href {http://dx.doi.org/10.1142/S0217751X08038135}
{\path{doi:10.1142/S0217751X08038135}}.

\bibitem{Machado:2007ea}
P.~F. Machado, F.~Saueressig, {On the renormalization group flow of
	f(R)-gravity}, Phys.Rev. D77 (2008) 124045.
\newblock \href {http://arxiv.org/abs/0712.0445} {\path{arXiv:0712.0445}},
\href {http://dx.doi.org/10.1103/PhysRevD.77.124045}
{\path{doi:10.1103/PhysRevD.77.124045}}.

\bibitem{Niedermaier:2009zz}
M.~R. Niedermaier, {Gravitational Fixed Points from Perturbation Theory}, Phys.
Rev. Lett. 103 (2009) 101303.
\newblock \href {http://dx.doi.org/10.1103/PhysRevLett.103.101303}
{\path{doi:10.1103/PhysRevLett.103.101303}}.

\bibitem{Benedetti:2009rx}
D.~Benedetti, P.~F. Machado, F.~Saueressig, {Asymptotic safety in
	higher-derivative gravity}, Mod. Phys. Lett. A24 (2009) 2233.
\newblock \href {http://arxiv.org/abs/0901.2984} {\path{arXiv:0901.2984}},
\href {http://dx.doi.org/10.1142/S0217732309031521}
{\path{doi:10.1142/S0217732309031521}}.

\bibitem{Benedetti:2009gn}
D.~Benedetti, P.~F. Machado, F.~Saueressig, {Taming perturbative divergences in
	asymptotically safe gravity}, Nucl. Phys. B824 (2010) 168.
\newblock \href {http://arxiv.org/abs/0902.4630} {\path{arXiv:0902.4630}},
\href {http://dx.doi.org/10.1016/j.nuclphysb.2009.08.023}
{\path{doi:10.1016/j.nuclphysb.2009.08.023}}.

\bibitem{Benedetti:2009iq}
D.~Benedetti, P.~F. Machado, F.~Saueressig, {Four-derivative interactions in
	asymptotically safe gravity}, AIP Conf. Proc. 1196 (2009) 44.
\newblock \href {http://arxiv.org/abs/0909.3265} {\path{arXiv:0909.3265}},
\href {http://dx.doi.org/10.1063/1.3284399} {\path{doi:10.1063/1.3284399}}.

\bibitem{Benedetti:2010nr}
D.~Benedetti, K.~Groh, P.~F. Machado, F.~Saueressig, {The Universal RG
	Machine}, JHEP 06 (2011) 079.
\newblock \href {http://arxiv.org/abs/1012.3081} {\path{arXiv:1012.3081}},
\href {http://dx.doi.org/10.1007/JHEP06(2011)079}
{\path{doi:10.1007/JHEP06(2011)079}}.

\bibitem{Rechenberger:2012pm}
S.~Rechenberger, F.~Saueressig, {The $R^2$ phase-diagram of QEG and its
	spectral dimension}, Phys. Rev. D86 (2012) 024018.
\newblock \href {http://arxiv.org/abs/1206.0657} {\path{arXiv:1206.0657}},
\href {http://dx.doi.org/10.1103/PhysRevD.86.024018}
{\path{doi:10.1103/PhysRevD.86.024018}}.

\bibitem{Ohta:2013uca}
N.~Ohta, R.~Percacci, {Higher Derivative Gravity and Asymptotic Safety in
	Diverse Dimensions}, Class. Quant. Grav. 31 (2014) 015024.
\newblock \href {http://arxiv.org/abs/1308.3398} {\path{arXiv:1308.3398}},
\href {http://dx.doi.org/10.1088/0264-9381/31/1/015024}
{\path{doi:10.1088/0264-9381/31/1/015024}}.

\bibitem{Falls:2013bv}
K.~Falls, D.~Litim, K.~Nikolakopoulos, C.~Rahmede, {A bootstrap towards
	asymptotic safety.} \href {http://arxiv.org/abs/1301.4191}
{\path{arXiv:1301.4191}}.

\bibitem{Falls:2014tra}
K.~Falls, D.~F. Litim, K.~Nikolakopoulos, C.~Rahmede, {Further evidence for
	asymptotic safety of quantum gravity}, Phys. Rev. D93 (2016) 104022.
\newblock \href {http://arxiv.org/abs/1410.4815} {\path{arXiv:1410.4815}},
\href {http://dx.doi.org/10.1103/PhysRevD.93.104022}
{\path{doi:10.1103/PhysRevD.93.104022}}.

\bibitem{Eichhorn:2015bna}
A.~Eichhorn, {The Renormalization Group flow of unimodular $f(R)$ gravity}, JHEP
04 (2015) 096.
\newblock \href {http://arxiv.org/abs/1501.05848} {\path{arXiv:1501.05848}},
\href {http://dx.doi.org/10.1007/JHEP04(2015)096}
{\path{doi:10.1007/JHEP04(2015)096}}.

\bibitem{Ohta:2015efa}
N.~Ohta, R.~Percacci, G.~P. Vacca, {Flow equation for $f(R)$ gravity and some
	of its exact solutions}, Phys. Rev. D92 (2015) 061501.
\newblock \href {http://arxiv.org/abs/1507.00968} {\path{arXiv:1507.00968}},
\href {http://dx.doi.org/10.1103/PhysRevD.92.061501}
{\path{doi:10.1103/PhysRevD.92.061501}}.

\bibitem{Falls:2016wsa}
K.~Falls, D.~F. Litim, K.~Nikolakopoulos, C.~Rahmede, {On de Sitter solutions
	in asymptotically safe $f(R)$ theories}, Class. Quant. Grav. 35 (2018)
135006.
\newblock \href {http://arxiv.org/abs/1607.04962} {\path{arXiv:1607.04962}},
\href {http://dx.doi.org/10.1088/1361-6382/aac440}
{\path{doi:10.1088/1361-6382/aac440}}.

\bibitem{Falls:2016msz}
K.~Falls, N.~Ohta, {Renormalization Group Equation for $f(R)$ gravity on
	hyperbolic spaces}, Phys. Rev. D94 (2016) 084005.
\newblock \href {http://arxiv.org/abs/1607.08460} {\path{arXiv:1607.08460}},
\href {http://dx.doi.org/10.1103/PhysRevD.94.084005}
{\path{doi:10.1103/PhysRevD.94.084005}}.

\bibitem{Christiansen:2016sjn}
N.~Christiansen, {Four-Derivative Quantum Gravity Beyond Perturbation
	Theory.} \href {http://arxiv.org/abs/1612.06223} {\path{arXiv:1612.06223}}.

\bibitem{Gonzalez-Martin:2017gza}
S.~Gonzalez-Martin, T.~R. Morris, Z.~H. Slade, {Asymptotic solutions in
	asymptotic safety}, Phys. Rev. D95 (2017) 106010.
\newblock \href {http://arxiv.org/abs/1704.08873} {\path{arXiv:1704.08873}},
\href {http://dx.doi.org/10.1103/PhysRevD.95.106010}
{\path{doi:10.1103/PhysRevD.95.106010}}.

\bibitem{Becker:2017tcx}
D.~Becker, C.~Ripken, F.~Saueressig, {On avoiding Ostrogradski instabilities
	within Asymptotic Safety,} JHEP 12 (2017) 121. \href {http://arxiv.org/abs/1709.09098}
{\path{arXiv:1709.09098}}.

\bibitem{Gies:2016con}
H.~Gies, B.~Knorr, S.~Lippoldt, F.~Saueressig, {Gravitational Two-Loop
	Counterterm Is Asymptotically Safe}, Phys. Rev. Lett. 116 (2016) 211302.
\newblock \href {http://arxiv.org/abs/1601.01800} {\path{arXiv:1601.01800}},
\href {http://dx.doi.org/10.1103/PhysRevLett.116.211302}
{\path{doi:10.1103/PhysRevLett.116.211302}}.

\bibitem{Reuter:2008qx}
M.~Reuter, H.~Weyer, {Conformal sector of Quantum Einstein Gravity in the local
	potential approximation: Non-Gaussian fixed point and a phase of unbroken
	diffeomorphism invariance}, Phys. Rev. D80 (2009) 025001.
\newblock \href {http://arxiv.org/abs/0804.1475} {\path{arXiv:0804.1475}},
\href {http://dx.doi.org/10.1103/PhysRevD.80.025001}
{\path{doi:10.1103/PhysRevD.80.025001}}.

\bibitem{Benedetti:2012dx}
D.~Benedetti, F.~Caravelli, {The Local potential approximation in quantum
	gravity}, JHEP 06 (2012) 017.
\newblock \href {http://arxiv.org/abs/1204.3541} {\path{arXiv:1204.3541}},
\href {http://dx.doi.org/10.1007/JHEP06(2012)017, 10.1007/JHEP10(2012)157}
{\path{doi:10.1007/JHEP06(2012)017, 10.1007/JHEP10(2012)157}}.

\bibitem{Demmel:2012ub}
M.~Demmel, F.~Saueressig, O.~Zanusso, {Fixed-Functionals of three-dimensional
	Quantum Einstein Gravity}, JHEP 11 (2012) 131.
\newblock \href {http://arxiv.org/abs/1208.2038} {\path{arXiv:1208.2038}},
\href {http://dx.doi.org/10.1007/JHEP11(2012)131}
{\path{doi:10.1007/JHEP11(2012)131}}.

\bibitem{Dietz:2012ic}
J.~A. Dietz, T.~R. Morris, {Asymptotic safety in the $f(R)$ approximation}, JHEP
01 (2013) 108.
\newblock \href {http://arxiv.org/abs/1211.0955} {\path{arXiv:1211.0955}},
\href {http://dx.doi.org/10.1007/JHEP01(2013)108}
{\path{doi:10.1007/JHEP01(2013)108}}.

\bibitem{Bridle:2013sra}
I.~H. Bridle, J.~A. Dietz, T.~R. Morris, {The local potential approximation in
	the background field formalism}, JHEP 03 (2014) 093.
\newblock \href {http://arxiv.org/abs/1312.2846} {\path{arXiv:1312.2846}},
\href {http://dx.doi.org/10.1007/JHEP03(2014)093}
{\path{doi:10.1007/JHEP03(2014)093}}.

\bibitem{Dietz:2013sba}
J.~A. Dietz, T.~R. Morris, {Redundant operators in the exact renormalisation
	group and in the f(R) approximation to asymptotic safety}, JHEP 07 (2013)
064.
\newblock \href {http://arxiv.org/abs/1306.1223} {\path{arXiv:1306.1223}},
\href {http://dx.doi.org/10.1007/JHEP07(2013)064}
{\path{doi:10.1007/JHEP07(2013)064}}.

\bibitem{Demmel:2014sga}
M.~Demmel, F.~Saueressig, O.~Zanusso, {RG flows of Quantum Einstein Gravity on
	maximally symmetric spaces}, JHEP 06 (2014) 026.
\newblock \href {http://arxiv.org/abs/1401.5495} {\path{arXiv:1401.5495}},
\href {http://dx.doi.org/10.1007/JHEP06(2014)026}
{\path{doi:10.1007/JHEP06(2014)026}}.

\bibitem{Demmel:2014hla}
M.~Demmel, F.~Saueressig, O.~Zanusso, {RG flows of Quantum Einstein Gravity in
	the linear-geometric approximation}, Annals Phys. 359 (2015) 141.
\newblock \href {http://arxiv.org/abs/1412.7207} {\path{arXiv:1412.7207}},
\href {http://dx.doi.org/10.1016/j.aop.2015.04.018}
{\path{doi:10.1016/j.aop.2015.04.018}}.

\bibitem{Demmel:2015oqa}
M.~Demmel, F.~Saueressig, O.~Zanusso, {A proper fixed functional for
	four-dimensional Quantum Einstein Gravity}, JHEP 08 (2015) 113.
\newblock \href {http://arxiv.org/abs/1504.07656} {\path{arXiv:1504.07656}},
\href {http://dx.doi.org/10.1007/JHEP08(2015)113}
{\path{doi:10.1007/JHEP08(2015)113}}.

\bibitem{Ohta:2015fcu}
N.~Ohta, R.~Percacci, G.~P. Vacca, {Renormalization Group Equation and scaling
	solutions for $f(R)$ gravity in exponential parametrization}, Eur. Phys. J.
C76 (2016) 46.
\newblock \href {http://arxiv.org/abs/1511.09393} {\path{arXiv:1511.09393}},
\href {http://dx.doi.org/10.1140/epjc/s10052-016-3895-1}
{\path{doi:10.1140/epjc/s10052-016-3895-1}}.

\bibitem{Labus:2016lkh}
P.~Labus, T.~R. Morris, Z.~H. Slade, {Background independence in a background
	dependent renormalization group}, Phys. Rev. D94 (2016) 024007.
\newblock \href {http://arxiv.org/abs/1603.04772} {\path{arXiv:1603.04772}},
\href {http://dx.doi.org/10.1103/PhysRevD.94.024007}
{\path{doi:10.1103/PhysRevD.94.024007}}.

\bibitem{Dietz:2016gzg}
J.~A. Dietz, T.~R. Morris, Z.~H. Slade, {Fixed point structure of the conformal
	factor field in quantum gravity}, Phys. Rev. D94 (2016) 124014.
\newblock \href {http://arxiv.org/abs/1605.07636} {\path{arXiv:1605.07636}},
\href {http://dx.doi.org/10.1103/PhysRevD.94.124014}
{\path{doi:10.1103/PhysRevD.94.124014}}.

\bibitem{Knorr:2017mhu}
B.~Knorr, {Infinite order quantum-gravitational correlations}, Class. Quant.
Grav. 35 (2018) 115005.
\newblock \href {http://arxiv.org/abs/1710.07055} {\path{arXiv:1710.07055}},
\href {http://dx.doi.org/10.1088/1361-6382/aabaa0}
{\path{doi:10.1088/1361-6382/aabaa0}}.

\bibitem{Falls:2017lst}
K.~Falls, C.~R. King, D.~F. Litim, K.~Nikolakopoulos, C.~Rahmede, {Asymptotic
	safety of quantum gravity beyond Ricci scalars}, Phys. Rev. D97 (2018)
086006.
\newblock \href {http://arxiv.org/abs/1801.00162} {\path{arXiv:1801.00162}},
\href {http://dx.doi.org/10.1103/PhysRevD.97.086006}
{\path{doi:10.1103/PhysRevD.97.086006}}.

\bibitem{Christiansen:2014raa}
N.~Christiansen, B.~Knorr, J.~M. Pawlowski, A.~Rodigast, {Global Flows in
	Quantum Gravity}, Phys. Rev. D93 (2016) 044036.
\newblock \href {http://arxiv.org/abs/1403.1232} {\path{arXiv:1403.1232}},
\href {http://dx.doi.org/10.1103/PhysRevD.93.044036}
{\path{doi:10.1103/PhysRevD.93.044036}}.

\bibitem{Meibohm:2015twa}
J.~Meibohm, J.~M. Pawlowski, M.~Reichert, {Asymptotic safety of gravity-matter
	systems}, Phys. Rev. D93 (2016) 084035.
\newblock \href {http://arxiv.org/abs/1510.07018} {\path{arXiv:1510.07018}},
\href {http://dx.doi.org/10.1103/PhysRevD.93.084035}
{\path{doi:10.1103/PhysRevD.93.084035}}.

\bibitem{Christiansen:2015rva}
N.~Christiansen, B.~Knorr, J.~Meibohm, J.~M. Pawlowski, M.~Reichert, {Local
	Quantum Gravity}, Phys. Rev. D92 (2015) 121501.
\newblock \href {http://arxiv.org/abs/1506.07016} {\path{arXiv:1506.07016}},
\href {http://dx.doi.org/10.1103/PhysRevD.92.121501}
{\path{doi:10.1103/PhysRevD.92.121501}}.

\bibitem{Denz:2016qks}
T.~Denz, J.~M. Pawlowski, M.~Reichert, {Towards apparent convergence in
	asymptotically safe quantum gravity}, Eur. Phys. J. C78 (2018) 336.
\newblock \href {http://arxiv.org/abs/1612.07315} {\path{arXiv:1612.07315}},
\href {http://dx.doi.org/10.1140/epjc/s10052-018-5806-0}
{\path{doi:10.1140/epjc/s10052-018-5806-0}}.

\bibitem{Christiansen:2017cxa}
N.~Christiansen, D.~F. Litim, J.~M. Pawlowski, M.~Reichert, {Asymptotic safety
	of gravity with matter}, Phys. Rev. D97 (2018) 106012.
\newblock \href {http://arxiv.org/abs/1710.04669} {\path{arXiv:1710.04669}},
\href {http://dx.doi.org/10.1103/PhysRevD.97.106012}
{\path{doi:10.1103/PhysRevD.97.106012}}.

\bibitem{Christiansen:2017bsy}
N.~Christiansen, K.~Falls, J.~M. Pawlowski, M.~Reichert, {Curvature dependence
	of quantum gravity}, Phys. Rev. D97~(4) (2018) 046007.
\newblock \href {http://arxiv.org/abs/1711.09259} {\path{arXiv:1711.09259}},
\href {http://dx.doi.org/10.1103/PhysRevD.97.046007}
{\path{doi:10.1103/PhysRevD.97.046007}}.

\bibitem{Eichhorn:2018akn}
A.~Eichhorn, P.~Labus, J.~M. Pawlowski, M.~Reichert, {Effective universality in
	quantum gravity}, SciPost Phys. 5~(4) (2018) 031.
\newblock \href {http://arxiv.org/abs/1804.00012} {\path{arXiv:1804.00012}},
\href {http://dx.doi.org/10.21468/SciPostPhys.5.4.031}
{\path{doi:10.21468/SciPostPhys.5.4.031}}.

\bibitem{Eichhorn:2018ydy}
A.~Eichhorn, S.~Lippoldt, J.~M. Pawlowski, M.~Reichert, M.~Schiffer, {How
	perturbative is quantum gravity?} Phys. Lett. B792 (2019) 310. \href {http://arxiv.org/abs/1810.02828}
{\path{arXiv:1810.02828}}, \href
{http://dx.doi.org/10.1016/j.physletb.2019.01.071}
{\path{doi:10.1016/j.physletb.2019.01.071}}.

\bibitem{Bosma:2019aiu}
L.~Bosma, B.~Knorr, F.~Saueressig, {Resolving Spacetime Singularities within
	Asymptotic Safety}, Phys. Rev. Lett. 123 (2019) 101301.
\newblock \href {http://arxiv.org/abs/1904.04845} {\path{arXiv:1904.04845}},
\href {http://dx.doi.org/10.1103/PhysRevLett.123.101301}
{\path{doi:10.1103/PhysRevLett.123.101301}}.

\bibitem{Knorr:2019atm}
B.~Knorr, C.~Ripken, F.~Saueressig, {Form Factors in Asymptotic Safety:
	conceptual ideas and computational toolbox}, Class. Quant. Grav. 36
(2019) 234001.
\newblock \href {http://arxiv.org/abs/1907.02903} {\path{arXiv:1907.02903}},
\href {http://dx.doi.org/10.1088/1361-6382/ab4a53}
{\path{doi:10.1088/1361-6382/ab4a53}}.

\bibitem{Dona:2013qba}
P.~Don\`a, A.~Eichhorn, R.~Percacci, {Matter matters in asymptotically safe
	quantum gravity}, Phys. Rev. D89 (2014) 084035.
\newblock \href {http://arxiv.org/abs/1311.2898} {\path{arXiv:1311.2898}},
\href {http://dx.doi.org/10.1103/PhysRevD.89.084035}
{\path{doi:10.1103/PhysRevD.89.084035}}.

\bibitem{Biemans:2017zca}
J.~Biemans, A.~Platania, F.~Saueressig, {Renormalization group fixed points of
	foliated gravity-matter systems}, JHEP 05 (2017) 093.
\newblock \href {http://arxiv.org/abs/1702.06539} {\path{arXiv:1702.06539}},
\href {http://dx.doi.org/10.1007/JHEP05(2017)093}
{\path{doi:10.1007/JHEP05(2017)093}}.

\bibitem{Eichhorn:2018whv}
A.~Eichhorn, A.~Held, {Mass difference for charged quarks from asymptotically
	safe quantum gravity}, Phys. Rev. Lett. 121 (2018) 151302.
\newblock \href {http://arxiv.org/abs/1803.04027} {\path{arXiv:1803.04027}},
\href {http://dx.doi.org/10.1103/PhysRevLett.121.151302}
{\path{doi:10.1103/PhysRevLett.121.151302}}.

\bibitem{Alkofer:2018fxj}
N.~Alkofer, F.~Saueressig, {Asymptotically safe $f(R)$-gravity coupled to
	matter I: the polynomial case}, Annals Phys. 396 (2018) 173.
\newblock \href {http://arxiv.org/abs/1802.00498} {\path{arXiv:1802.00498}},
\href {http://dx.doi.org/10.1016/j.aop.2018.07.017}
{\path{doi:10.1016/j.aop.2018.07.017}}.

\bibitem{Pawlowski:2018ixd}
J.~M. Pawlowski, M.~Reichert, C.~Wetterich, M.~Yamada, {Higgs scalar potential
	in asymptotically safe quantum gravity,}  Phys. Rev. D99 (2019) 086010. \href
{http://arxiv.org/abs/1811.11706} {\path{arXiv:1811.11706}}.

\bibitem{deBrito:2019epw}
G.~P. De~Brito, Y.~Hamada, A.~D. Pereira, M.~Yamada, {On the impact of Majorana
	masses in gravity-matter systems}, JHEP 08 (2019) 142.
\newblock \href {http://arxiv.org/abs/1905.11114} {\path{arXiv:1905.11114}},
\href {http://dx.doi.org/10.1007/JHEP08(2019)142}
{\path{doi:10.1007/JHEP08(2019)142}}.

\bibitem{Burger:2019upn}
B.~B{\"u}rger, J.~M. Pawlowski, M.~Reichert, B.-J. Schaefer, {Curvature
	dependence of quantum gravity with scalars.} \href
{http://arxiv.org/abs/1912.01624} {\path{arXiv:1912.01624}}.

\bibitem{Eichhorn:2018yfc}
A.~Eichhorn, {An asymptotically safe guide to quantum gravity and matter,} Front. Astron. Space Sci. 5 (2019) 47. \href
{http://arxiv.org/abs/1810.07615} {\path{arXiv:1810.07615}}.

\bibitem{Morris:2018zgy}
T.~R. Morris, R.~Percacci, {Trace anomaly and infrared cutoffs}, Phys. Rev.
D99 (2019) 105007.
\newblock \href {http://arxiv.org/abs/1810.09824} {\path{arXiv:1810.09824}},
\href {http://dx.doi.org/10.1103/PhysRevD.99.105007}
{\path{doi:10.1103/PhysRevD.99.105007}}.

\bibitem{Baldazzi:2018mtl}
A.~Baldazzi, R.~Percacci, V.~Skrinjar, {Wicked metrics}, Class. Quant. Grav.
36 (2019) 105008.
\newblock \href {http://arxiv.org/abs/1811.03369} {\path{arXiv:1811.03369}},
\href {http://dx.doi.org/10.1088/1361-6382/ab187d}
{\path{doi:10.1088/1361-6382/ab187d}}.

\bibitem{Baldazzi:2019kim}
A.~Baldazzi, R.~Percacci, V.~Skrinjar, {Quantum fields without Wick rotation},
Symmetry 11 (2019) 373.
\newblock \href {http://arxiv.org/abs/1901.01891} {\path{arXiv:1901.01891}},
\href {http://dx.doi.org/10.3390/sym11030373}
{\path{doi:10.3390/sym11030373}}.

\bibitem{Maas:2019eux}
A.~Maas, {The Fr{\"o}hlich-Morchio-Strocchi mechanism and quantum gravity.} \href
{http://arxiv.org/abs/1908.02140} {\path{arXiv:1908.02140}}.

\bibitem{Ambjorn:2012jv}
J.~Ambjorn, A.~Goerlich, J.~Jurkiewicz, R.~Loll, {Nonperturbative Quantum
	Gravity}, Phys. Rept. 519 (2012) 127.
\newblock \href {http://arxiv.org/abs/1203.3591} {\path{arXiv:1203.3591}},
\href {http://dx.doi.org/10.1016/j.physrep.2012.03.007}
{\path{doi:10.1016/j.physrep.2012.03.007}}.

\bibitem{Loll:2019rdj}
R.~Loll, {Quantum Gravity from Causal Dynamical Triangulations: A Review,} Class. Quant. Grav. 37 (2020) 013002. \href
{http://arxiv.org/abs/1905.08669} {\path{arXiv:1905.08669}}.

\bibitem{Ambjorn:2011cg}
J.~Ambjorn, S.~Jordan, J.~Jurkiewicz, R.~Loll, {A Second-order phase transition
	in CDT}, Phys. Rev. Lett. 107 (2011) 211303.
\newblock \href {http://arxiv.org/abs/1108.3932} {\path{arXiv:1108.3932}},
\href {http://dx.doi.org/10.1103/PhysRevLett.107.211303}
{\path{doi:10.1103/PhysRevLett.107.211303}}.

\bibitem{Ambjorn:2012ij}
J.~Ambjorn, S.~Jordan, J.~Jurkiewicz, R.~Loll, {Second- and First-Order Phase
	Transitions in CDT}, Phys. Rev. D85 (2012) 124044.
\newblock \href {http://arxiv.org/abs/1205.1229} {\path{arXiv:1205.1229}},
\href {http://dx.doi.org/10.1103/PhysRevD.85.124044}
{\path{doi:10.1103/PhysRevD.85.124044}}.

\bibitem{Ambjorn:2019pkp}
J.~Ambjorn, D.~Coumbe, J.~Gizbert-Studnicki, A.~G{\"o}rlich, J.~Jurkiewicz,
{Critical Phenomena in Causal Dynamical Triangulations}, Class. Quant. Grav.
36 (2019) 224001.
\newblock \href {http://arxiv.org/abs/1904.05755} {\path{arXiv:1904.05755}},
\href {http://dx.doi.org/10.1088/1361-6382/ab4184}
{\path{doi:10.1088/1361-6382/ab4184}}.

\bibitem{Ambjorn:2019lrm}
J.~Ambjorn, J.~Gizbert-Studnicki, A.~G{\"o}rlich, J.~Jurkiewicz, D.~Nemeth,
{Towards an UV fixed point in CDT gravity}, JHEP 07 (2019) 166.
\newblock \href {http://arxiv.org/abs/1906.04557} {\path{arXiv:1906.04557}},
\href {http://dx.doi.org/10.1007/JHEP07(2019)166}
{\path{doi:10.1007/JHEP07(2019)166}}.

\bibitem{Manrique:2011jc}
E.~Manrique, S.~Rechenberger, F.~Saueressig, {Asymptotically Safe Lorentzian
	Gravity}, Phys. Rev. Lett. 106 (2011) 251302.
\newblock \href {http://arxiv.org/abs/1102.5012} {\path{arXiv:1102.5012}},
\href {http://dx.doi.org/10.1103/PhysRevLett.106.251302}
{\path{doi:10.1103/PhysRevLett.106.251302}}.

\bibitem{Rechenberger:2012dt}
S.~Rechenberger, F.~Saueressig, {A functional renormalization group equation
	for foliated spacetimes}, JHEP 03 (2013) 010.
\newblock \href {http://arxiv.org/abs/1212.5114} {\path{arXiv:1212.5114}},
\href {http://dx.doi.org/10.1007/JHEP03(2013)010}
{\path{doi:10.1007/JHEP03(2013)010}}.

\bibitem{Biemans:2016rvp}
J.~Biemans, A.~Platania, F.~Saueressig, {Quantum gravity on foliated
	spacetimes: Asymptotically safe and sound}, Phys. Rev. D95 (2017) 086013.
\newblock \href {http://arxiv.org/abs/1609.04813} {\path{arXiv:1609.04813}},
\href {http://dx.doi.org/10.1103/PhysRevD.95.086013}
{\path{doi:10.1103/PhysRevD.95.086013}}.

\bibitem{Houthoff:2017oam}
W.~B. Houthoff, A.~Kurov, F.~Saueressig, {Impact of topology in foliated
	Quantum Einstein Gravity}, Eur. Phys. J. C77 (2017) 491.
\newblock \href {http://arxiv.org/abs/1705.01848} {\path{arXiv:1705.01848}},
\href {http://dx.doi.org/10.1140/epjc/s10052-017-5046-8}
{\path{doi:10.1140/epjc/s10052-017-5046-8}}.

\bibitem{Knorr:2018fdu}
B.~Knorr, {Lorentz symmetry is relevant}, Phys. Lett. B792 (2019) 142.
\newblock \href {http://arxiv.org/abs/1810.07971} {\path{arXiv:1810.07971}},
\href {http://dx.doi.org/10.1016/j.physletb.2019.01.070}
{\path{doi:10.1016/j.physletb.2019.01.070}}.

\bibitem{Eichhorn:2019ybe}
A.~Eichhorn, A.~Platania, M.~Schiffer, {Lorentz invariance violations in the
	interplay of quantum gravity with matter.} \href
{http://arxiv.org/abs/1911.10066} {\path{arXiv:1911.10066}}.

\bibitem{Ambjorn:2005db}
J.~Ambjorn, J.~Jurkiewicz, R.~Loll, {Spectral dimension of the universe}, Phys.
Rev. Lett. 95 (2005) 171301.
\newblock \href {http://arxiv.org/abs/hep-th/0505113}
{\path{arXiv:hep-th/0505113}}, \href
{http://dx.doi.org/10.1103/PhysRevLett.95.171301}
{\path{doi:10.1103/PhysRevLett.95.171301}}.

\bibitem{Lauscher:2005qz}
O.~Lauscher, M.~Reuter, {Fractal spacetime structure in asymptotically safe
	gravity}, JHEP 10 (2005) 050.
\newblock \href {http://arxiv.org/abs/hep-th/0508202}
{\path{arXiv:hep-th/0508202}}, \href
{http://dx.doi.org/10.1088/1126-6708/2005/10/050}
{\path{doi:10.1088/1126-6708/2005/10/050}}.

\bibitem{Reuter:2012xf}
M.~Reuter, F.~Saueressig, {Asymptotic Safety, Fractals, and Cosmology}, Lect.
Notes Phys. 863 (2013) 185.
\newblock \href {http://arxiv.org/abs/1205.5431} {\path{arXiv:1205.5431}},
\href {http://dx.doi.org/10.1007/978-3-642-33036-0_8}
{\path{doi:10.1007/978-3-642-33036-0_8}}.

\bibitem{Carlip:2019onx}
S.~Carlip, {Dimension and Dimensional Reduction in Quantum Gravity}, Universe 5
(2019) 83.
\newblock \href {http://arxiv.org/abs/1904.04379} {\path{arXiv:1904.04379}},
\href {http://dx.doi.org/10.3390/universe5030083}
{\path{doi:10.3390/universe5030083}}.

\bibitem{Pagani:2017tdr}
C.~Pagani, H.~Sonoda, {Products of composite operators in the exact
	renormalization group formalism}, PTEP 2018 (2018) 023B02.
\newblock \href {http://arxiv.org/abs/1707.09138} {\path{arXiv:1707.09138}},
\href {http://dx.doi.org/10.1093/ptep/ptx189}
{\path{doi:10.1093/ptep/ptx189}}.

\bibitem{Becker:2018quq}
M.~Becker, C.~Pagani, {Geometric operators in the asymptotic safety scenario
	for quantum gravity}, Phys. Rev. D99 (2019) 066002.
\newblock \href {http://arxiv.org/abs/1810.11816} {\path{arXiv:1810.11816}},
\href {http://dx.doi.org/10.1103/PhysRevD.99.066002}
{\path{doi:10.1103/PhysRevD.99.066002}}.

\bibitem{Becker:2019tlf}
M.~Becker, C.~Pagani, {Geometric Operators in the Einstein-Hilbert Truncation},
Universe 5~(3) (2019) 75.
\newblock \href {http://dx.doi.org/10.3390/universe5030075}
{\path{doi:10.3390/universe5030075}}.

\bibitem{Becker:2019fhi}
M.~Becker, C.~Pagani, O.~Zanusso, {Fractal geometry of higher derivative
	gravity.} \href{http://arxiv.org/abs/1911.02415}{\path{arXiv:1911.02415}}.

\bibitem{Klitgaard:2017ebu}
N.~Klitgaard, R.~Loll, {Introducing Quantum Ricci Curvature}, Phys. Rev.
D97 (2018) 046008.
\newblock \href {http://arxiv.org/abs/1712.08847} {\path{arXiv:1712.08847}},
\href {http://dx.doi.org/10.1103/PhysRevD.97.046008}
{\path{doi:10.1103/PhysRevD.97.046008}}.

\bibitem{Klitgaard:2018snm}
N.~Klitgaard, R.~Loll, {Implementing quantum Ricci curvature}, Phys. Rev.
D97 (2018) 106017.
\newblock \href {http://arxiv.org/abs/1802.10524} {\path{arXiv:1802.10524}},
\href {http://dx.doi.org/10.1103/PhysRevD.97.106017}
{\path{doi:10.1103/PhysRevD.97.106017}}.

\bibitem{Reuter:1993kw}
M.~Reuter, C.~Wetterich, {Effective average action for gauge theories and exact
	evolution equations}, Nucl. Phys. B417 (1994) 181.
\newblock \href {http://dx.doi.org/10.1016/0550-3213(94)90543-6}
{\path{doi:10.1016/0550-3213(94)90543-6}}.

\bibitem{Demmel:2013myx}
M.~Demmel, F.~Saueressig, O.~Zanusso,
\href{http://inspirehep.net/record/1217855/files/arXiv:1302.1312.pdf}{{Fixed
		Functionals in Asymptotically Safe Gravity}}, in: {Proceedings, 13th Marcel
	Grossmann Meeting on Recent Developments in Theoretical and Experimental
	General Relativity, Astrophysics, and Relativistic Field Theories (MG13)} 
(2015) 2227.
\newblock \href {http://arxiv.org/abs/1302.1312} {\path{arXiv:1302.1312}},
\href {http://dx.doi.org/10.1142/9789814623995_0404}
{\path{doi:10.1142/9789814623995_0404}}.
%\newline\urlprefix\url{http://inspirehep.net/record/1217855/files/arXiv:1302.1312.pdf}

\bibitem{Dietz:2015owa}
J.~A. Dietz, T.~R. Morris, {Background independent exact renormalization group
	for conformally reduced gravity}, JHEP 04 (2015) 118.
\newblock \href {http://arxiv.org/abs/1502.07396} {\path{arXiv:1502.07396}},
\href {http://dx.doi.org/10.1007/JHEP04(2015)118}
{\path{doi:10.1007/JHEP04(2015)118}}.

\bibitem{deBrito:2018jxt}
G.~P. De~Brito, N.~Ohta, A.~D. Pereira, A.~A. Tomaz, M.~Yamada, {Asymptotic
	safety and field parametrization dependence in the $f(R)$ truncation}, Phys.
Rev. D98 (2018) 026027.
\newblock \href {http://arxiv.org/abs/1805.09656} {\path{arXiv:1805.09656}},
\href {http://dx.doi.org/10.1103/PhysRevD.98.026027}
{\path{doi:10.1103/PhysRevD.98.026027}}.

\bibitem{Ohta:2018sze}
N.~Ohta, R.~Percacci, A.~D. Pereira, {$f(R, R_{\mu\nu}^2)$ at one loop}, Phys.
Rev. D97 (2018) 104039.
\newblock \href {http://arxiv.org/abs/1804.01608} {\path{arXiv:1804.01608}},
\href {http://dx.doi.org/10.1103/PhysRevD.97.104039}
{\path{doi:10.1103/PhysRevD.97.104039}}.

\bibitem{Falls:2018ylp}
K.~G. Falls, D.~F. Litim, J.~Schr{\"o}der, {Aspects of asymptotic safety for
	quantum gravity.} \href {http://arxiv.org/abs/1810.08550}
{\path{arXiv:1810.08550}}.

\bibitem{Alkofer:2018baq}
N.~Alkofer, {Asymptotically safe $f(R)$-gravity coupled to matter II: Global
	solutions}, Phys. Lett. B789 (2019) 480--487.
\newblock \href {http://arxiv.org/abs/1809.06162} {\path{arXiv:1809.06162}},
\href {http://dx.doi.org/10.1016/j.physletb.2018.12.061}
{\path{doi:10.1016/j.physletb.2018.12.061}}.

\bibitem{Pawlowski:2005xe}
J.~M. Pawlowski, {Aspects of the functional renormalisation group}, Annals
Phys. 322 (2007) 2831.
\newblock \href {http://arxiv.org/abs/hep-th/0512261}
{\path{arXiv:hep-th/0512261}}, \href
{http://dx.doi.org/10.1016/j.aop.2007.01.007}
{\path{doi:10.1016/j.aop.2007.01.007}}.

\bibitem{Igarashi:2009tj}
Y.~Igarashi, K.~Itoh, H.~Sonoda, {Realization of Symmetry in the ERG Approach
	to Quantum Field Theory}, Prog. Theor. Phys. Suppl. 181 (2010) 1.
\newblock \href {http://arxiv.org/abs/0909.0327} {\path{arXiv:0909.0327}},
\href {http://dx.doi.org/10.1143/PTPS.181.1} {\path{doi:10.1143/PTPS.181.1}}.

\bibitem{Pagani:2016pad}
C.~Pagani, {Note on scaling arguments in the effective average action
	formalism}, Phys. Rev. D94 (2016) 045001.
\newblock \href {http://arxiv.org/abs/1603.07250} {\path{arXiv:1603.07250}},
\href {http://dx.doi.org/10.1103/PhysRevD.94.045001}
{\path{doi:10.1103/PhysRevD.94.045001}}.

\bibitem{Ellwanger:1994iz}
U.~Ellwanger, {Flow equations and BRS invariance for Yang-Mills theories},
Phys. Lett. B335 (1994) 364.
\newblock \href {http://arxiv.org/abs/hep-th/9402077}
{\path{arXiv:hep-th/9402077}}, \href
{http://dx.doi.org/10.1016/0370-2693(94)90365-4}
{\path{doi:10.1016/0370-2693(94)90365-4}}.

\bibitem{DAttanasio:1996tzp}
M.~D'Attanasio, T.~R. Morris, {Gauge invariance, the quantum action principle,
	and the renormalization group}, Phys. Lett. B378 (1996) 213.
\newblock \href {http://arxiv.org/abs/hep-th/9602156}
{\path{arXiv:hep-th/9602156}}, \href
{http://dx.doi.org/10.1016/0370-2693(96)00411-X}
{\path{doi:10.1016/0370-2693(96)00411-X}}.

\bibitem{Litim:1998qi}
D.~F. Litim, J.~M. Pawlowski, {Flow equations for Yang-Mills theories in
	general axial gauges}, Phys. Lett. B435 (1998) 181.
\newblock \href {http://arxiv.org/abs/hep-th/9802064}
{\path{arXiv:hep-th/9802064}}, \href
{http://dx.doi.org/10.1016/S0370-2693(98)00761-8}
{\path{doi:10.1016/S0370-2693(98)00761-8}}.

\bibitem{Codello:2008vh}
A.~Codello, R.~Percacci, C.~Rahmede, {Investigating the Ultraviolet Properties
	of Gravity with a Wilsonian Renormalization Group Equation}, Annals Phys. 324
(2009) 414.
\newblock \href {http://arxiv.org/abs/0805.2909} {\path{arXiv:0805.2909}},
\href {http://dx.doi.org/10.1016/j.aop.2008.08.008}
{\path{doi:10.1016/j.aop.2008.08.008}}.

\bibitem{Litim:2000ci}
D.~F. Litim, {Optimization of the exact renormalization group}, Phys. Lett. B486
(2000) 92.
\newblock \href {http://arxiv.org/abs/hep-th/0005245}
{\path{arXiv:hep-th/0005245}}, \href
{http://dx.doi.org/10.1016/S0370-2693(00)00748-6}
{\path{doi:10.1016/S0370-2693(00)00748-6}}.

\bibitem{Litim:2001up}
D.~F. Litim, {Optimized renormalization group flows}, Phys. Rev. D64 (2001)
105007.
\newblock \href {http://arxiv.org/abs/hep-th/0103195}
{\path{arXiv:hep-th/0103195}}, \href
{http://dx.doi.org/10.1103/PhysRevD.64.105007}
{\path{doi:10.1103/PhysRevD.64.105007}}.

\bibitem{Reuter:2011ah}
M.~Reuter, F.~Saueressig, {Fractal space-times under the microscope: A
	Renormalization Group view on Monte Carlo data}, JHEP 12 (2011) 012.
\newblock \href {http://arxiv.org/abs/1110.5224} {\path{arXiv:1110.5224}},
\href {http://dx.doi.org/10.1007/JHEP12(2011)012}
{\path{doi:10.1007/JHEP12(2011)012}}.

\bibitem{Ambjorn:2008wc}
J.~Ambjorn, A.~Gorlich, J.~Jurkiewicz, R.~Loll, {The Nonperturbative Quantum de
	Sitter Universe}, Phys. Rev. D78 (2008) 063544.
\newblock \href {http://arxiv.org/abs/0807.4481} {\path{arXiv:0807.4481}},
\href {http://dx.doi.org/10.1103/PhysRevD.78.063544}
{\path{doi:10.1103/PhysRevD.78.063544}}.

\bibitem{Ambjorn:2016fbd}
J.~Ambjørn, Z.~Drogosz, J.~Gizbert-Studnicki, A.~Görlich, J.~Jurkiewicz,
D.~Nemeth, {Impact of topology in causal dynamical triangulations quantum
	gravity}, Phys. Rev. D94 (2016) 044010.
\newblock \href {http://arxiv.org/abs/1604.08786} {\path{arXiv:1604.08786}},
\href {http://dx.doi.org/10.1103/PhysRevD.94.044010}
{\path{doi:10.1103/PhysRevD.94.044010}}.

\bibitem{Knorr:2018kog}
B.~Knorr, F.~Saueressig, {Towards reconstructing the quantum effective action
	of gravity}, Phys. Rev. Lett. 121 (2018) 161304.
\newblock \href {http://arxiv.org/abs/1804.03846} {\path{arXiv:1804.03846}},
\href {http://dx.doi.org/10.1103/PhysRevLett.121.161304}
{\path{doi:10.1103/PhysRevLett.121.161304}}.

\bibitem{Vassilevich:2003xt}
D.~V. Vassilevich, {Heat kernel expansion: User's manual}, Phys. Rept. 388
(2003) 279.
\newblock \href {http://arxiv.org/abs/hep-th/0306138}
{\path{arXiv:hep-th/0306138}}, \href
{http://dx.doi.org/10.1016/j.physrep.2003.09.002}
{\path{doi:10.1016/j.physrep.2003.09.002}}.

\bibitem{York:1973ia}
J.~W. York, Jr., {Conformatlly invariant orthogonal decomposition of symmetric
	tensors on Riemannian manifolds and the initial value problem of general
	relativity}, J. Math. Phys. 14 (1973) 456.
\newblock \href {http://dx.doi.org/10.1063/1.1666338}
{\path{doi:10.1063/1.1666338}}.


\end{thebibliography}
\end{document}